\documentclass[12pt]{article}
\usepackage{latexsym}
\usepackage{amsthm}
\usepackage{amsmath}
\usepackage[dvips]{graphicx}
\usepackage{amssymb}
\usepackage{multirow}

\begin{document}
\baselineskip 0.6cm

\def\simgt{\mathrel{\lower2.5pt\vbox{\lineskip=0pt\baselineskip=0pt
           \hbox{$>$}\hbox{$\sim$}}}}
\def\simlt{\mathrel{\lower2.5pt\vbox{\lineskip=0pt\baselineskip=0pt
           \hbox{$<$}\hbox{$\sim$}}}}

\begin{titlepage}

\begin{flushright}
UCB-PTH-10/18 \\
\end{flushright}

\vskip 2.0cm

\begin{center}

{\LARGE \bf 
Origins of Hidden Sector Dark Matter II: \\ \vspace{0.25cm} Collider Physics
}

\vskip 1.0cm

{\large Clifford Cheung$^{1,2}$, Gilly Elor$^{1,2}$, Lawrence J. Hall$^{1,2,3}$ and Piyush Kumar$^{1,2,4}$}

\vskip 0.4cm

$^{1}${\it Berkeley Center for Theoretical Physics, Department of Physics, \\
     University of California, Berkeley, CA 94720} \\
$^{2}${\it Theoretical Physics Group, Lawrence Berkeley National Laboratory,
     Berkeley, CA 94720} \\

$^{3}${\it Institute for the Physics and Mathematics of the Universe, \\ University of Tokyo, Kashiwa 277-8568, Japan}\\


$^{4}${\it Department of Physics,  Columbia University, New York, NY 10027}

\abstract{

We consider a broad class of supersymmetric theories in which dark matter (DM) is the lightest superpartner (LSP) of a hidden sector that couples very weakly to visible sector fields.  
Portal interactions connecting visible and hidden sectors mediate the decay of the lightest observable superpartner (LOSP) into the LSP, allowing the LHC to function as a spectacular probe of the origin of hidden sector DM.  
As shown in a companion paper, this general two-sector framework allows only for a handful of DM production mechanisms, each of which maps to a distinctive window in lifetimes and cross-sections for the LOSP.  In the present work we perform a systematic collider study of LOSP candidates and portal interactions, and for each case evaluate the prospects for successfully reconstructing the origin of DM at the LHC.  If, for instance, DM arises from Freeze-Out and Decay, this may be verified if the LOSP is a bino or right-handed slepton decaying to the LSP through a variety of portal interactions, and with an annihilation cross-section within a narrow range.  On the other hand, the Freeze-In mechanism may be verified for a complimentary set of LOSP candidates, and within a narrow range of LOSP lifetimes.  
In all cases, the LOSP is relatively long-lived on collider time scales, leading to events with displaced vertices.  Furthermore,  scenarios with a charged or colored LOSP are particularly promising.
}

\end{center}
\end{titlepage}

\def\simgt{\mathrel{\lower2.5pt\vbox{\lineskip=0pt\baselineskip=0pt
           \hbox{$>$}\hbox{$\sim$}}}}
\def\simlt{\mathrel{\lower2.5pt\vbox{\lineskip=0pt\baselineskip=0pt
           \hbox{$<$}\hbox{$\sim$}}}}

\renewcommand{\l}{\langle}
\renewcommand{\r}{\rangle}
\newcommand{\be}{\begin{eqnarray}}
\newcommand{\ee}{\end{eqnarray}}

\newcommand{\dd}[2]{\frac{\partial #1}{\partial #2}}
\newcommand{\NN}{\mathcal{N}}
\newcommand{\LL}{\mathcal{L}}
\newcommand{\MM}{\mathcal{M}}
\newcommand{\ZZ}{\mathcal{Z}}
\newcommand{\WW}{\mathcal{W}}

\section{Introduction}

In the standard picture of weakly interacting massive particle (WIMP) dark matter (DM), a weak scale DM particle is in thermal equilibrium with a bath of visible sector particles at some high temperature.  As the Universe expands and the temperature drops, the DM undergoes thermal Freeze-Out (FO) \cite{Kolb:1988aj}, yielding a cold DM abundance.  As is well-known, the thermal DM relic abundance is uniquely determined by the DM annihilation cross-section, a quantity which can in principle be accessed at colliders.  Hence, the WIMP paradigm offers the remarkable prospect of reconstructing the cosmological origin of DM at the LHC \cite{Baltz:2006fm}.

In recent years, however, developments in top-down approaches to particle physics have begun to indicate that the above WIMP paradigm corresponds to a rather restricted class of scenarios in which there is a single sector (the visible sector) which DM is part of and with which it shares sizeable interactions. It is possible to envisage a much broader framework in which there exists more than one sector with the hidden sector containing its own particles and interactions with masses and couplings not too different from the visible sector. This framework throws open many exciting possibilities for DM and subsumes the standard WIMP paradigm as a special case.    

In this work, our focus will be the possibility that DM resides in a hidden sector that couples very weakly to visible sector fields but contains its own set of particles and interactions.  In this scenario the visible and  hidden sectors are {\it separately} in thermal equilibrium and exhibit a rich cosmology which has been studied extensively in a companion paper \cite{Cheung:2010gj}.  The present work is a systematic collider study evaluating the prospects of identifying the origin of hidden sector DM within this broad class of theories at the LHC.

\begin{figure}[t] 
  \center{
\includegraphics[scale=1.5]{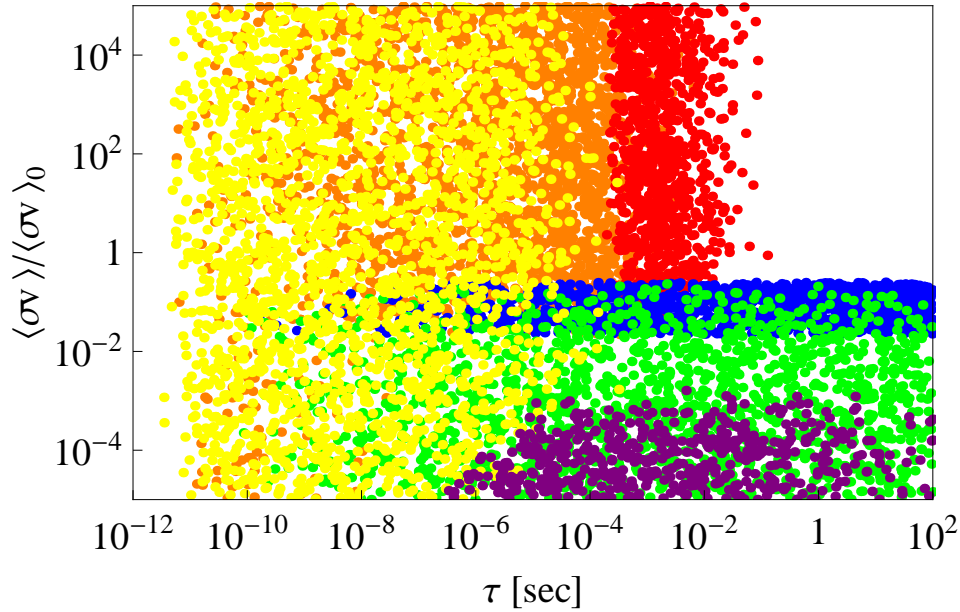}
} 
\caption{\footnotesize{As shown in \cite{cosmopaper}, hidden sector DM can arise through a handful of production mechanisms, each corresponding to a distinct window in the $\tau - \langle \sigma v\rangle$ plane.  Here \{FO\&D, FO\&D$_{\rm r}$, FO\&D$_{\rm a}$, FI, FI$_{\rm r}$, FI$_{\rm a}$\} are denoted by \{blue, green, purple, red, orange, yellow\}.   The re-annihilation mechanisms \{FO\&D$_{\rm r}, $ FI$_{\rm r}$\} occur when a sufficiently large abundance of $X'$ particles are produced in FO\&D or FI such that $X'$ annihilation starts up again.  Meanwhile, the mechanisms  \{FO\&D$_{\rm a}, $ FI$_{\rm a}$\} result when the FO\&D and FI processes generate a particle anti-particle  asymmetry for $X'$.   Every point corresponds to the observed DM abundance of $\Omega h^2 =0.11$ and we have inclusively scanned over the remaining parameters relevant to the cosmological evolution, including $10 \, \mbox{GeV} < m < 1 \, \mbox{TeV}$ and $1/20 < m'/m < 1/2$.  See \cite{cosmopaper} for details.   }}
\label{fig:introplot}
\end{figure}

To begin, we assume that DM is charged under a stabilizing symmetry shared by the visible and hidden sectors.  We denote the lightest visible and hidden sector particles charged under this symmetry by $X$ and $X'$, and take their masses $m > m'$ to be broadly of order the weak scale.  By construction, $X'$ is the DM particle.  Furthermore, we assume that portal interactions between the visible and hidden sectors mediate the decay process
\be
X &\rightarrow& X' + \ldots,
\label{eq:XtoX'}
\ee
where the ellipses denote what are typically visible decay products.     As shown in \cite{cosmopaper}, our setup allows only for a handful of mechanisms which can account for the present day abundance of DM.   Our primary focus will be on the following DM production mechanisms: 
\begin{itemize}
\item {\bf Freeze-Out and Decay (FO\&D).}  $X$ undergoes FO and then decays out of equilibrium, yielding an abundance of $X'$.  The final abundance of $X'$ goes as \cite{cosmopaper}
\be
\Omega &\propto& 
\frac{m'}{m\langle \sigma v\rangle}.
\label{eq:FODeq}
\ee
\item {\bf Freeze-In (FI). } $X$ decays while still in thermal equilibrium with the visible sector, yielding an abundance of $X'$. The final abundance of $X'$ goes as \cite{cosmopaper,Hall:2009bx}
\be
\Omega & \propto&  
\frac{m'}{m^2 \tau}.
\label{eq:FIeq}
\ee
\end{itemize}
The above formulas imply that the origin of DM is correlated with a characteristic window in the parameter space defined by $\tau$, the lifetime of $X$, and $\langle \sigma v\rangle$, the thermally averaged annihilation cross-section of $X$ (see Figure \ref{fig:introplot}). 
Remarkably, FO\&D and FI depend solely on parameters which might some day be measured at the LHC!   In particular, since $X$ is a visible sector field there is a chance that it will be produced directly in high energy collisions.  If this is the case then its mass, lifetime, and annihilation cross-section can in principle be measured.  At the same time, the visible decay products in Eq.~\ref{eq:XtoX'} may be analyzed for an event by event reconstruction of the decay process in order to determine the mass of $X'$.
Interestingly, if we require that the visible and hidden sectors do not thermally equilibrate at the weak scale, then this implies
\be
\tau &>& 10^{-13} \textrm{ s} \;\; \left( \frac{100 \textrm{ GeV}}{m}\right)^2  \left( \frac{100}{g'_*(T \simeq m)/g_X} \right),
\label{eq:tau}
\ee
where $g'_* (g_X)$ are the number of spin degrees of freedom of the hidden sector ($X$).
This implies that the broad class of theories studied in this paper will typically exhibit displaced vertices from the decay of $X$.  The aim of the present work is to determine a systematic blueprint for how the origin of DM might be reconstructed at the LHC. The connection between a very weakly coupled dark sector and displaced vertices at colliders as 
well as the possibility of reconstructing the cosmology from such measurements, was first pointed out in 
\cite{Chang:2009sv}. 

To this end, we consider a concrete supersymmetric realization of the scenario described above.  Indeed, supersymmetry offers the ideal stabilizing symmetry for DM, i.e.~R-parity, while gravity mediated supersymmetry breaking provides a compelling theoretical explanation for the existence of weak scale states in both the visible and hidden sectors.  In the language of supersymmetry, $X$ is then the lightest observable sector superpartner (LOSP) while $X'$ is the lightest superpartner (LSP).

In the single sector MSSM, the neutral superpartners $\tilde{b}, \tilde{w}, \tilde{h}$ and $\tilde{\nu}$ are all candidates for DM.  However, for masses of interest the FO yields are too high for $\tilde{b}$ and too low for $\tilde{w}, \tilde{h}$ and $\tilde{\nu}$.  Successful DM typically requires the LSP to be a careful mixture of these states or for other states to have accidental degeneracies \cite{ADG, King:2006tf}.  However, in two sector cosmologies $\tilde{b}$ becomes an ideal candidate for the LOSP that gives DM via FO\&D, while $\tilde{w}, \tilde{h}$ and $\tilde{\nu}$ are ideal LOSP candidates for DM from FI.  Furthermore any charged or colored LOSP allows DM to be dominated by FI, while the right-handed slepton also allows FO\&D.
\begin{table}[t]
\begin{center}
\begin{tabular}{||c||c|c||}
\hline \hline
 &  Freeze-Out and Decay (FO\&D) &  Freeze-In (FI) \\
\hline \hline
 LOSP  & $\tilde{\chi}_0, \tilde{\ell} $ & $\tilde q$, $\tilde \ell$, $\tilde \nu$, $\tilde g$, $\tilde \chi_0$, $\tilde \chi_\pm$ \\
\hline
 Operators & $\mathcal{O}X'$  & $H_u H_d X'$, $B^\alpha X_\alpha' $  \\
\hline
Observables & $ m, m',\langle \sigma v \rangle$ & $m, m', \tau$   \\
\hline 
Range & $ 10^{-27}\textrm{ cm}^3/\textrm{s}< \langle \sigma v \rangle <10^{-26}\textrm{ cm}^3/\textrm{s}$ & $ 10^{-4}\textrm{ s} < \tau < 10^{-1}\textrm{ s}$   \\
\hline 
Predicted Relation & $\frac{m' \langle \sigma v \rangle_0}{m \langle \sigma v\rangle} =  1 $ &  $\frac{m'}{m \tau} \left( \frac{100 \textrm{ GeV}}{m}\right) = 25\text{ s}^{-1}$  \\
\hline 
\hline
\end{tabular} 
\end{center}
\caption{\footnotesize{The origin of DM may be fully reconstructed for a specific set of LOSP candididates and portal operators.  If the designated observables are measured, we should discover they lie in the ranges listed above, and satisfy the predicted relations given schematically in Eqs.~\ref{eq:FODeq} and \ref{eq:FIeq} and precisely in the last row of the table.  Here $\langle \sigma v\rangle_0 = 3\times 10^{-26}\textrm{ cm}^3/\textrm{s}$ and $\mathcal{O}$ denotes an operator of dimension $\leq 4$ comprised of visible sector fields.}}
\label{tab:introtab}
\end{table}

 A priori, the identity of the LOSP is unknown, as is the nature of its couplings to the LSP.  Scanning over all possible LOSP candidates and portal operators, we obtain Table \ref{tab:introtab}, which summarizes the circumstances under which FO\&D and FI might be fully reconstructed at the LHC.  For each mechanism of DM production one requires a specific combination of LOSP candidates and operators.  Furthermore, in order to measure the observables designated in Table \ref{tab:introtab}, it is necessary to specify the particular decay processes which are relevant for each choice of LOSP and portal operator (see Tables \ref{tab:FOD1}, \ref{tab:FOD2}, and \ref{tab:FI}).  As we will see, the nature of the LOSP, i.e.~whether it is charged or colored, will have a significant impact on whether these observables can truly be probed at the LHC. However, in the event that these observables are successfully measured, we may discover that they lie in the ranges and satisfy the predicted relations specified in Table \ref{tab:introtab}.  If so, the LHC would provide a spectacular and unequivocal verification of the cosmological origin of hidden sector DM.

It is of course possible that $\tau$ and $\langle \sigma v\rangle$ will be measured but found to lie outside the expected ranges for FO\&D and FI depicted in Figure \ref{fig:introplot}.  Such an observation could be consistent with DM arising from variants of FO\&D and FI, namely  \{FO\&D$_{\rm r}$, FO\&D$_{\rm a}$, FI$_{\rm r}$, FI$_{\rm a}$\}, discussed in significant detail in \cite{cosmopaper}.  As shown in \cite{cosmopaper} these mechanisms are strongly dependent on parameters which are inaccessible to colliders\footnote{In particular, FO\&D$_{\rm r}$ and FI$_{\rm r}$ depend on the annihilation cross-section of $X'$, while FO\&D$_{\rm a}$ and FI$_{\rm a}$ depend on the CP violation in processes that connect the sectors.}, so a smoking gun verification of these cosmologies at the LHC is unlikely.  At the same time, these variant DM production mechanisms correspond to a particular set of collider signatures which depend on the portal interaction. For FO\&D$_{\rm r}$ and FI$_{\rm r}$, the relevant signatures include those associated with FO\&D and FI, which are contained in Tables  \ref{tab:FOD1}, \ref{tab:FOD2}, and \ref{tab:FI}.  Furthermore, FO\&D$_{\rm r}$ and FI$_{\rm r}$ also include those signatures shown in Tables \ref{tab:REsig} and \ref{tab:ROsig}.  These signature tables also apply to FO\&D$_{\rm a}$ and FI$_{\rm a}$.  Similar results were found in an analysis of displaced collider signatures associated with asymmetric DM \cite{Chang:2009sv}. 

Our paper is structured as follows.  In Section \ref{sec:twosec} we present a brief summary of our supersymmetric two-sector setup.  We go on to consider all possible portal interactions between the visible and hidden sector in Section \ref{sec:opanalysis}.  In Sections \ref{sec:FOD} and \ref{sec:FI} we discuss FO\&D and FI and their prospects for reconstruction at the LHC.  We go on in Section \ref{detect} to consider how the nature of the LOSP, namely whether it is charged or colored, effects our ability to measure the required observables for reconstructing the cosmology.  Finally, in Section \ref{sec:con} we conclude and discuss the signatures associated with re-annihilated and asymmetric versions of the FO\&D and FI mechanisms.

\section{Two-Sector Setup}
\label{sec:twosec}

  We assume throughout that DM is stabilized by an exact R-parity shared by the visible and hidden sectors.  We denote all superfields in upper-case and all component fields in lower-case.  Furthermore, all R-parity odd component fields will appear with a tilde, so for instance the quark, squark, and quark superfield will be denoted by $q$, $\tilde q$, and $Q$.  In this notation, $X$ and $X'$ are the superfields containing the LOSP, $\tilde x$, and the LSP, $\tilde x'$.    

Our discussion will be free from theoretical prejudices on supersymmetry breaking and so the identity of the LOSP will be unconstrained by UV considerations:
\be
X \in \{Q,U,D,L,E,H_u,H_d,B^\alpha,W^\alpha,G^\alpha \}.
\ee  
Our primary focus will be the interplay between cosmology and collider phenomenology in the cases of FO\&D and FI.  In order to reconstruct our cosmological history at the LHC it will be necessary to extract several important observables from colliders, as summarized in Table \ref{tab:introtab}.  For FO\&D, we must measure or infer the annihilation cross-section and mass of the LOSP, as well as the mass of the LSP.  For FI, we must measure or infer the partial width of every MSSM field into the LSP, as well as the masses of the LOSP and LSP.  In both cases, we are interested in the prospect of measuring the LSP mass from the long-lived decay process,
\be 
\tilde x \rightarrow \tilde x' + \textrm{SM}
\ee
Along these lines, a subset of LOSP candidates have been studied extensively in the literature.  
For instance, the so-called superWIMP is a remarkable proposal in which $\tilde x = \tilde \ell^\pm$ is a charged slepton and  $\tilde x' = \tilde G$ is the gravitino \cite{Feng:2003uy,Feng:2004zu}.  In these scenarios, the long-lived decay $\tilde \ell^\pm \rightarrow \ell^\pm \tilde G$ can be probed in colliders to reconstruct the cosmological history of the universe.   Others have considered alternative LOSP candidates, such as the bino, as well as other possible LSP candidates, such as an axino \cite{Covi:1999ty} or goldstino \cite{Cheung:2010mc,Cheung:2010qf}.
\begin{figure}[t] 
\begin{center}  \includegraphics[scale=.8]{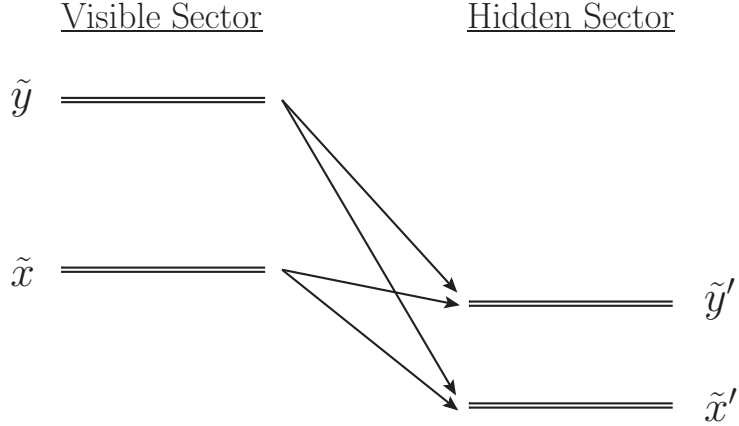}\end{center}
\caption{\footnotesize{Allowed decays among visible and hidden sector states.  Here $\tilde x$ and $\tilde x'$ denote the LOSP and LSP, and $\tilde y$ and $\tilde y'$ collectively denote the spectator fields.  Only $\tilde x \rightarrow (\tilde x', \tilde y')+\textrm{SM}$ can be measured at colliders.  The process $\tilde x \rightarrow \tilde x' + \textrm{SM}$ must be visible in order to measure the LSP mass.}}
\label{fig:alldecays}
\end{figure}
Certainly, the existing literature has been largely focused on measuring decays of the LOSP to the LSP.  However, the philosophy of this work is that the hidden sector may be as rich in spectrum and dynamics as our own, and so the visible and hidden sectors actually contain many more fields than just $X$ and $X'$.  

First of all, there of course exist numerous additional visible sector superfields which we will denote by $Y$, whose R-parity odd component we denote by $\tilde y$.  Here $\tilde y$ could be a heavy squark or gluino, for example.  Secondly, the hidden sector may contain additional superfields which we collectively denote by $Y'$, whose R-parity odd component we denote by $\tilde y'$.  Given our working premise that visible and hidden sector fields interact weakly, then there in general exist operators which couple $X,Y$ to $X',Y'$, yielding the decay modes (see Figure \ref{fig:alldecays})
\be
\tilde x &\rightarrow& \tilde x' +\textrm{SM}\\
\tilde x &\rightarrow& \tilde y' +\textrm{SM}\\
\tilde y &\rightarrow& \tilde x' +\textrm{SM}\\
\tilde y &\rightarrow& \tilde y' +\textrm{SM}
\ee
At the LHC, MSSM cascades will terminate at $\tilde x$, which in turn decays via $\tilde x \rightarrow (\tilde x',\tilde y') + \textrm{SM}$.  As long as there is a sufficiently large branching ratio of the LOSP to the LSP via visible decay modes, then the LSP mass may be measured, which is essential for reconstructing the cosmological history of FO\&D and FI.  

What are the constraints on the gravitino in this two sector cosmology?  First, the gravitino must be heavier than the hidden sector LSP, otherwise the gravitino itself would be the LSP.  This requires a large scale of supersymmetry breaking, as in gravity or anomaly mediation.  If the gravitino is lighter than the LOSP, then it has no effect on our cosmology: LOSP decays to gravitinos are highly suppressed compared to decays to the LSP, and any gravitinos produced at very high temperatures will decay to hidden sector states and not affect nucleosynthesis.  On the other hand, if the gravitino is heavier than the LOSP a standard gravitino problem emerges, with decays to visible sector superpartners limiting the reheat temperature to about $10^6$ GeV.



\section{Operator Analysis}
\label{sec:opanalysis}

We assume that the LSP is a gauge singlet and catalog all dimension $d\leq 5$ gauge-invariant operators that connect it to visible sector fields.  These operators in turn dictate the allowed decay modes of the LOSP.  We organize our analysis in terms of the symmetry structure of each of these operators, and summarize all of our results for FO\&D and FI in tables \ref{tab:FOD1} + \ref{tab:FOD2} and  \ref{tab:FI}, respectively.  We will provide a detailed explanation of these tables in Sections \ref{FOD-signals} and \ref{FI-signals}.

$X'$ couples to the visible sector via $d\leq 5$ operators of the form
\be
   \mathcal{O} X'^{(\dagger)}
\label{eq:OXp}
\ee
where ${\cal O}$ is a gauge invariant, dimension $\leq 4$ operator comprised of visible sector fields and $\mathcal{O}X'^{(\dagger)}$ appears in the Kahler potential or superpotential depending on the holomorphy properties of this operator.  Note that $\mathcal{O}$ need  {\it not}  directly contain the LOSP, $X$!

As we will see, whether $\mathcal{O} X'^{(\dagger)}$ appears in the Kahler potential or the superpotential will often dictate whether the portal to the hidden sector is renormalizable or not.  We will define a dimensionless coupling $\lambda$ which characterizes the strength of the portal interaction in Eq.~(\ref{eq:OXp}).  If $d=4$, then $\lambda$ is simply the coefficient of the marginal portal interaction.  However, for $d>4$ higher dimension operators, we can define 
\be 
\lambda \equiv (m/M_*)^{d-4}
\label{eq:lambdadef}
\ee
where $M_*$ is the scale of the higher dimension operator.

Assuming that the visible sector is the MSSM, we can catalog  all operators $\mathcal{O}$ according to their transformation properties under R-parity and R-symmetry, as shown in Table \ref{table:optable}.  There we have defined the operators
\begin{table}
\begin{center}
\begin{tabular}{||c||c|c|c||c||c|c|c|c||} \hline \hline
  & ${\cal O}_K$ & ${\cal O}_W$ & $H_u H_d$ & $B^\alpha$ & $LH_u$ & $LH_d^\dagger$ & $LLE, QLD$& $UDD$\\ \hline \hline
R-parity & $+$ & $+$ & $+$ & $-$ & $-$ & $-$ & $-$ & $-$\\ \hline 
R-charge & $0$ & $2$ & $R_1$ & $1$ & $R_2$ & $R_2-R_1$ & $2+R_2-R_1$ & $R_3$ \\ \hline\hline
\end{tabular} 
\end{center}
\caption{\footnotesize{Table of R-parity and R-charge assignments for various visible sector operators.  Here $\mathcal{O}_K$ and $\mathcal{O}_W$ are defined in Eq.~(\ref{eq:ops}).}}
\label{table:optable}
\end{table}
\be
\mathcal{O}_{K} &=& \{Q^\dagger Q, U^\dagger U, D^\dagger D, L^\dagger L, E^\dagger E, H_u^\dagger H_u, H_d^\dagger H_d \} 
\nonumber \\ 
\mathcal{O}_{W} &=&\{B^\alpha B_\alpha,W^\alpha W_\alpha,G^\alpha G_\alpha,  QH_u U, QH_d D, LH_d E\} ,
\label{eq:ops}
\ee
which correspond to operators which appear in the Kahler and superpotential of the MSSM.  Note that $\mathcal{O}_{K}$, $\mathcal{O}_{W}$, and $B^\alpha$ have fixed R-charge because they are present in the MSSM Lagrangian.  In contrast, the remaining operators  have unspecified R-charge because they are R-parity odd and are thus absent from the MSSM Lagrangian.  Furthermore, $H_u H_d$ has unspecified R-charge because we wish to be agnostic about the origin of the $\mu$ parameter.  

Two of the above operators are particularly noteworthy as viable portals between the visible and hidden sectors.  First, coupling through $B^\alpha X'_\alpha$ induces a gauge kinetic mixing between $U(1)$ vector superfields in the visible and hidden sector fields \cite{Holdom:1985ag}.  This ``Bino Portal'' has been studied extensively in terms of hidden sector phenomenology \cite{Glashow:1985ud,Pospelov:2007mp,ArkaniHamed:2008qp,Baumgart:2009tn,Cheung:2009qd,Cheung:2009su,Walker:2009ei}, and we provide a brief review in Appendix \ref{app:bino}.  Second of all, there exists an operator $H_u H_d X'$ which, after EWSB, induces a mass mixing between the LSP and the higgsinos.  We briefly review this ``Higgs Portal'' theory in Appendix \ref{higgsportal-model} .  The Bino and Higgs Portals are distinct from the operator portals shown in Eq.~\ref{table:optable} in that they induce relevant kinetic and mass mixings, respectively, between $\tilde x'$ and MSSM neutralinos.

The R-parity and R-charge of $X'$ dictates which subset of operators can function as a portal between the visible and hidden sectors.  For each group of operators in Table \ref{table:optable}, we have catalogued the associated decay modes of the LOSP, given various choices for the LOSP.  Our results for FO\&D are summarized in Tables \ref{tab:FOD1} and \ref{tab:FOD2}, and for FI in Table \ref{tab:FI}.  We assume all operators in a group have comparable coefficients, so that some groups have many possible decay modes.  Because we are concerned with the measurement of the LSP mass via $\tilde x \rightarrow \tilde x' + \textrm{SM}$, we have only listed the leading  decay mode, as well as any subdominant (three-body) decay modes which are lepton rich and thus more useful in reconstructing the event.  For this reason, we do not include subdominant decay modes involving only quarks or Higgs bosons.  Moreover, we include the parametric size of each branching ratio as a function of masses, mixing angles, couplings, and three-body phase space factors. In doing that, we have omitted the coefficients of the various connector operators (which can be different by ${\cal O}(1)$) and have assumed that $m \sim v \sim m_{3/2}$ are parametrically the same, so the dependences on these three quantities is just shown as a dependence on $m$. 

There is an important difference between Tables \ref{tab:FOD1} and \ref{tab:FOD2} for FO\&D on the one hand and Table \ref{tab:FI} for FI on the other, which highlights the  complementarity of these two cosmological scenarios. In particular, FO\&D (Tables \ref{tab:FOD1} and \ref{tab:FOD2}) can be achieved for many different operators but only for two LOSP candidates: a bino-like neutralino or a right-handed slepton.  
In contrast, FI (Table \ref{tab:FI}) can be achieved for many different LOSP candidates, but only for a restricted subset of operators.  The reasons for the asymmetry are discussed in detail in Sections \ref{sec:FOD} and \ref{sec:FI}.

Before concluding our operator analysis, we wish to address a number of related issues.  First of all, there is the matter of supersymmetry breaking operators, which we have ignored thus far.  In particular, any supersymmetry respecting coupling between the visible and hidden sector can be dressed with supersymmetry breaking spurions yielding a supersymmetry breaking operator.  For example, if ${\cal O}X'^{(\dagger)}$ is allowed by symmetries, then there can also be operators of the form  $f(\Phi,\Phi^\dagger){\cal O} X'^{(\dagger)} $, where $f$ is an arbitrary complex function of a supersymmetry breaking spurion, $\Phi=\theta^2 m_{3/2}$.  In general, these supersymmetry breaking operators provide new decay modes for the LOSP which can affect cosmology.  We will discuss the effects of these operators on FO\&D and FI in Sections \ref{sec:FOD} and \ref{sec:FI}. 


Secondly, we note that our discussion has been purposefully general with respect to R-parity and R-charge assignments in the operators listed in Table \ref{table:optable}.  However, particular choices of the ultraviolet physics can be used to fix $R_{i}$.  For instance, if neutrino masses are generated above the supersymmetry breaking scale, then $(LH_u)^2$ is present in the superpotential and so $R_2 =1$.  If the $\mu$ term is generated by the Giudice-Masiero mechanism, which is natural for gravity mediated supersymmetry breaking, then $R_1 =0$, assuming that all supersymmetry breaking spurions have R-charge zero.  With both of these assumptions, all of the R-charges are fixed except $R_3$.

\section{Freeze-Out and Decay (FO\&D)}
\label{sec:FOD}

In the case of FO\&D, $\tilde x$ undergoes thermal FO and then decays late into the hidden sector.  Throughout, we assume that the annihilation cross-section and mass of $\tilde x$ can be measured at colliders, and refer the reader to detailed analyses on this topic \cite{Baltz:2006fm}.  Once these quantities are known, the thermal relic abundance of $\tilde x$ is straightforwardly inferred.  The mass of the LSP can be measured in colliders via $\tilde x \rightarrow \tilde x' +\textrm{SM}$.   Still, what is the effect of the additional decay processes, $\tilde x \rightarrow \tilde y' + \textrm{SM}$ and $\tilde y \rightarrow (\tilde x', \tilde y') + \textrm{SM}$? 

\begin{figure}
\begin{center}  \includegraphics[scale=.8]{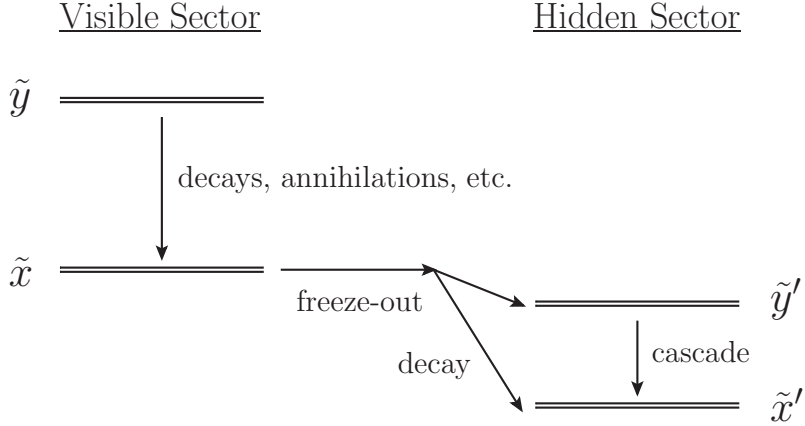}\end{center}
\caption{\footnotesize{Depiction of the cosmological history of FO\&D. As the temperature drops below their masses, heavy visible superparters, $\tilde y$, are depleted to a negligible level by decay and annihilation.   As the temperature falls below the mass of the LOSP, $\tilde x$, they annihilate and freeze-out; these relic particles decay much later into the hidden sector.  Consequently, direct decays of $\tilde y$ to the hidden sector are irrelevant.
Also, due to R-parity all decays of $\tilde x$ ultimately yield the LSP, $\tilde x'$.    To reconstuct this cosmology we must measure the annihilation cross-section and mass of $\tilde x$ and the mass of $\tilde x'$. }}
\label{fig:FOD}
\end{figure}

\subsection{Spectator Fields in Freeze-Out and Decay}

First of all, since $\tilde y$ has no FO abundance, all of its decays are irrelevant to the cosmology of FO\&D.  This is fortunate because the partial width of $\tilde y$ into the hidden sector is too tiny to be measured directly at colliders.  Second of all, while $\tilde x$ can decay to $\tilde y'$, these hidden sector particles  ultimately cascade down to $\tilde y'$ due to R-parity conservation.  Consequently, every $\tilde x$ decay yields precisely an odd number of  $\tilde x'$ particles.  In the vast majority of cases, a single $\tilde x'$ is produced in each $\tilde x$ decay, but in some instances three or more may be produced.  We will elaborate on this possibility later on when we discuss FO\&D via the Bino ($B^\alpha X'_\alpha$) and Higgs ($H_u H_d X'$) Portals.  As long as there is a non-negligible branching fraction of $\tilde x$ to visibly decay into $\tilde x'$, then the LSP mass can be measured\footnote{In principle, the branching fraction to the LSP can be fairly small, since for instance in $\tilde \tau \rightarrow \tau \tilde x'$, the endpoint in the tau-stau invariant mass distribution indicates the mass of the {\it lightest} decay product.}.  Thus, we have argued that the decay processes in Figure \ref{fig:FOD} are largely irrelevant except for $\tilde x\rightarrow  \tilde x' + \textrm{SM}$, which is necessary to measure the LSP mass.  

\subsection{Theories of Freeze-Out and Decay}

Since FO\&D can be verified at the LHC quite model independently, the only constraint 
on the LOSP candidate for FO\&D is that the FO abundance is sufficiently large. As we will explain, this is possible 
if the LOSP is a bino-like neutralino or a right-handed slepton. 

In particular, since yield of LSP particles from FO\&D is precisely equal to the abundance of $X$ which arises from visible sector FO, the final energy density of DM arising from FO\&D will be $m'/m$ smaller than the FO abundance that would have been produced in a single sector theory with the same annihilation cross-section.  This implies that obtaining the observed relic abundance of DM from FO\&D requires the LOSP to overproduce by a factor of $m/m'$. Hence the 
bino is an ideal candidate for the LOSP since in the MSSM a bino LOSP with 
$ m_{\tilde{b}} < 100 \textrm{ GeV}$ already yields the correct relic abundance 
$\Omega h^2 \sim 0.11$. Since the bino can 
annihilate via exchange of a right-handed slepton, the bino cross-section depends on the slepton mass in addition   
to the bino mass (see Eq.~(2) of \cite{ArkaniHamed:2006mb}). Fixing the 
dark matter abundance and requiring a bino LOSP sets an upper  
bound on the bino mass. For the bino to overproduce the resulting bound is $m_{\tilde{b}} < 250\textrm{ GeV}$ 
for $m^\prime/m_{\tilde{b}} > 1/20$. 

Similarly for the case of the right-handed slepton, a diagram involving t-channel bino 
exchange results in bino mass dependent cross-section:

\begin{equation}\label{sigma-l}
\sigma v_{\tilde{l}_R} \sim \frac{4 \pi \alpha^2}{m_{\tilde{l}_R}^2} + 
\frac{16 \pi \alpha^2 m_{\tilde{b}}^2}{\cos^4{\theta_w} 
\left(m_{\tilde{l}_R}^2 + m_{\tilde{b}}^2 \right)^2}
\end{equation}
Requiring that the right-handed slepton be the LOSP, and that $\tilde{b}$ is not closely degenerate with $\tilde{l}_{R}$ results in the lower 
bound $m_{\tilde{l}_R} > 700 \textrm{ GeV}$ for $m^\prime /m_{\tilde{l}_R} < 1/2$. This makes $\tilde{l}_R$ a less attractive LOSP candidate then $\tilde{b}$. Other MSSM LOSP candidates would need to be even heavier.

Thus, the primary constraint on theories of FO\&D are on the identity of the LOSP.  On the other hand, the nature of portal interactions are essentially irrelevant, as long as the decay products from the decay of $\tilde x$ are sufficient to reconstruct the LSP mass.  As we will see in the following sections, and in Tables \ref{tab:FOD1} and \ref{tab:FOD2}, a broad range of portal interactions allow for lepton rich decay channels which are promising for constructing the LSP mass.  Next, let 
us consider the collider signatures of neutralino and slepton LOSP in turn.


\begin{table}[t]
\centering
\small\addtolength{\tabcolsep}{-5pt}
\renewcommand{\arraystretch}{1.8}
\begin{center}
\begin{tabular}{|c||c|c|c|c|c|}
\hline
\hline
\multicolumn{6}{|c|}{FO$\&$D} \\
\hline
\hline
&
& \multicolumn{2}{|c|}{$\tilde\chi_{0}$} 
& \multicolumn{2}{|c|}{$\tilde{\ell^\pm}$} \\ \cline{3-6}

 Operator & Charges ($X'$) & Decay & $k$ & Decay & $k$ \\
\hline
\hline
 \multirow{2}{*}{$\cal O$$_{K}X^\prime$}
& $(+,0)$ & $\tilde\chi_{0} \rightarrow \ell^+ \ell^- \tilde{x}^{\prime}$ 
& $\frac{1}{(4 \pi)^2} g_{\tilde\chi\tilde\ell\ell}^2 \frac{m^4}{m_{\tilde{l}}^4}$ 
& $\tilde{\ell}^\pm \rightarrow \ell^\pm \tilde{x}^{\prime}$ & l \\ 
&  & $\tilde\chi_{0}\rightarrow (h^0,Z)\tilde{x}^{\prime}$ 
& $\theta_{\tilde\chi \tilde h}^2, \theta_{\tilde\chi \tilde h}^2 g_{2}^2 $  &  &  \\
\hline
 \multirow{2}{*}{$\cal O$$_{W} X^\prime$} 
& $(+,0)$ & $\chi_{0} \rightarrow (\gamma,Z)\tilde{x}^{\prime}$ 
& $\theta_{\tilde \chi\tilde{b}}^2,\theta_{\tilde\chi \tilde{w}}^2$ 
& $\tilde{\ell}^\pm \rightarrow \ell^\pm  (\gamma,Z) \tilde{x}^{\prime}$ 
& $\frac{1}{(4 \pi)^2} m^2 (\frac{g_{1\ell}^2}{m_{\tilde b}^2},\frac{g_{2}^2}{m_{\tilde w}^2} )$ \\ 
&  & $\tilde\chi_{0} \rightarrow \ell^+ \ell^- \tilde{x}^{\prime}$ 
& $\frac{1}{(4 \pi)^2} g_{\tilde\chi\tilde\ell\ell}^2 \frac{m^4}{m_{\tilde{l}}^4}$  
& $\tilde{\ell}^\pm \rightarrow \ell^\pm \tilde{x}^{\prime}$ & 1 \\
\hline
 \multirow{3}{*}{$ H_u H_d X^\prime \left(X^{\prime\dagger}\right) $} 
& & $\tilde\chi_0 \rightarrow (h^0,Z)\tilde{x}^{\prime}$ 
& $\theta_{\tilde \chi \tilde h}^2, \theta_{\tilde\chi \tilde h}^2 g_{2}^2$ 
& $\tilde{\ell}^\pm \rightarrow \ell^\pm \tilde{x}^{\prime}$ 
& $g_{\tilde h \tilde\ell\ell}^2 $ \\ 
& $(+,2-R_1) $ or $(+,R_1)$  & $\tilde \chi_0 \rightarrow y^{\prime} \tilde{y}^{\prime}$ & $\theta_{\tilde{\chi}\tilde{h}}^2\lambda^{\prime 2}$ &  &  \\ 
&  & $\tilde \chi_{0} \rightarrow \ell^+\ell^- \tilde{x}^{\prime}$ & $\frac{1}{(4 \pi)^2} g_{\tilde\chi\tilde\ell\ell}^2 g_{1\ell}^2 \frac{m^4}{m_{\tilde{l}}^4}$ & & \\
\hline
\end{tabular}
\end{center}
\caption{\footnotesize{Decay modes and decay rates for LOSP candidates $\tilde\chi_{0}$ 
and $\tilde{\ell}$, relevant for FO\&D, with $X'$ R-parity even.  The second column lists the R-parity and R-charge of $X'$.  Here $g_{\tilde\chi \tilde\ell \ell}\equiv\theta_{\tilde\chi \tilde{b}} g_{1\ell} +\theta_{\tilde \chi \tilde w} g_{2}+\theta_{\tilde\chi \tilde{h}} \lambda_{\ell}$ is the effective coupling between $\tilde \chi_0$ and $\tilde\ell \ell$ and $g_{\tilde h \tilde\ell \ell}\equiv \lambda_\ell + v(g_{1\ell}g_{1h} /m_{\tilde b} + \theta_{\tilde \ell \tilde\ell_L} g_{2}^2/m_{\tilde w})$ is the effective coupling between $\tilde h$ and $\tilde\ell \ell$, with mass mixing calculated using an insertion approximation.  Here $k$ characterizes the size of the partial width for each process, as defined in Eq.~\ref{eq:kdef}.
}}
\label{tab:FOD1}
\end{table}

\begin{table}[t]
\centering
\small\addtolength{\tabcolsep}{-5pt}
\renewcommand{\arraystretch}{1.8}
\begin{center}
\begin{tabular}{|c||c|c|c|c|c|}
\hline
\hline
\multicolumn{6}{|c|}{FO$\&$D} \\
\hline
\hline
&
& \multicolumn{2}{|c|}{$\tilde\chi_{0}$} 
& \multicolumn{2}{|c|}{$\tilde{\ell^\pm}$} \\ \cline{3-6}

 Operator & Charges ($X'$) & Decay & $k$ & Decay & $k$ \\ 
\hline
\hline
 \multirow{3}{*}{$ B^\alpha X^{\prime}_\alpha $} 
& & $\tilde \chi_{0} \rightarrow y^{\prime} \tilde{y}^{\prime}$ 
& $\theta_{\tilde{\chi}\tilde{b}}^2 g^{\prime 2}$ & $\tilde{\ell}^\pm \rightarrow \ell^\pm \tilde{x}^{\prime}$ & $g_{1\ell}^2$ \\
& $(-,1)$ & $\tilde \chi_{0} \rightarrow \ell^+\ell^- \tilde{x}^{\prime}$ 
& $\frac{1}{(4 \pi)^2} g_{\tilde\chi\tilde\ell\ell}^2 g_{1\ell}^2 \frac{m^4}{m_{\tilde{l}}^4}$  & & \\
& & $\tilde\chi_0 \rightarrow (h^0,Z)\tilde{x}^{\prime}$ & $\theta_{\tilde \chi \tilde h}^2, \theta_{\tilde\chi \tilde h}^2 g_{2}^2$& & \\
\hline
 \multirow{2}{*}{$ LH_u X^\prime $} 
& $(-,2-R_2)$ & $\tilde\chi_0 \rightarrow \nu \tilde{x}^{\prime}$ & $\theta_{\tilde\chi \tilde{h}}^2$ 
& $\tilde{\ell}^\pm \rightarrow \ell^\pm \nu \tilde{x}^{\prime}$ 
& $\frac{1}{(4 \pi)^2} g_{\tilde h \tilde\ell\ell}^2 \frac{m^2}{m_{\tilde{h}}^2}$ \\ 
&  & $\tilde\chi_0 \rightarrow \ell^\pm (h^\mp,W^\mp)\tilde x'$ & $\frac{1}{(4\pi)^2} g_{2}^2 \frac{m^2}{m_{\tilde h}^2} ( \theta_{\tilde\chi \tilde w}^2,\theta_{\tilde\chi \tilde h}^2)$  & $\tilde{\ell}^{\pm} \rightarrow (h^{\pm},W^{\pm})\tilde{x}^{\prime}$ & $\theta_{\tilde\ell \tilde \ell_L}^2 (1,g_{2}^2)$ \\

\hline
$ LH_u X^{\prime\dagger} $ & $(-,R_2)$ & '' 
& ''
& '' & '' \\
\hline
$ LH_d^\dagger X^{\prime\dagger}$ & $(-, R_2-R_1)$ 
& '' & '' & '' & '' \\

$ LH_d^\dagger X^\prime $ & $(-, R_1-R_2)$ & '' & '' & '' & '' \\

\hline
 $ LLEX^\prime, QLDX^\prime $ & $(-,R_1-R_2)$ 
& $\tilde\chi_0 \rightarrow l^+ l^- \nu \tilde{x}^{\prime}$ 
& $\frac{1}{(4 \pi)^4} g_{\tilde\chi \tilde\ell\ell}^2 \frac{m^4}{m_{\tilde{l}}^4}$ 
& $\tilde{\ell}^\pm \rightarrow \ell^\pm \nu  \tilde{x}^{\prime}$ & $\frac{1}{(4 \pi)^2}$ \\
\hline
\end{tabular}
\end{center}
\caption{\footnotesize{Decay modes and decay rates for LOSP candidates $\tilde\chi_{0}$ 
and $\tilde{\ell}$, relevant for FO\&D, with $X'$ R-parity odd.  Here $k$, $g_{\tilde\chi \tilde\ell \ell}$, and $g_{\tilde h \tilde\ell \ell}$ are defined as in Table \ref{tab:FOD1}.}}
\label{tab:FOD2}
\end{table}

\subsection{Collider Signatures of Freeze-Out and Decay}
\label{FOD-signals}

In this section, we consider associated collider signatures for FO\&D for the right-handed slepton and bino-like neutralino.  Tables \ref{tab:FOD1} and \ref{tab:FOD2} provide an extensive summary of the decay processes relevant for reconstructing the cosmological history. The structure and notation in these tables require a bit of explanation. 

 Each row corresponds to a possible choice for the portal operator coupling the visible and hidden sectors.  Here Table  \ref{tab:FOD1} (\ref{tab:FOD2}) corresponds to the R-parity even (odd) $X'$.  Along each row in each table, we have presented the (R-parity, R-charge) assignments for $X'$ required for the corresponding portal interaction.  Also, along each row is the information characterizing the collider signatures for each choice of LOSP, which in the case of FO\&D can be $\tilde \chi_0$ or $\tilde \ell^\pm$.  For each LOSP, we list the leading decay channel of the LOSP, as well as subdominant decay channels which contain leptons or Higgs and gauge bosons, which may decay leptonically.  These lepton--rich channels are more promising for event reconstruction.  Note that $y'$ and $\tilde y'$ denoted in the neutralino decay via the Higgs Portal operator are used to denote hidden sector particles which are effectively invisible in the collider.  

Each decay process is associated with a partial width
\be
\Gamma({\tilde x \rightarrow \tilde{x}'+ \textrm{SM}}) &=& \left( \frac{1}{8\pi} \lambda^2 m \right)  k({\tilde x \rightarrow \tilde{x}'+ \textrm{SM}})
\label{eq:kdef}
\ee
where $\lambda$ is the coefficient of the portal operator if it is marginal, and otherwise is defined as in Eq.~\ref{eq:lambdadef}.  Here the dimensionless parameter $k$ is presented for each decay process in Tables \ref{tab:FOD1} and \ref{tab:FOD2}.  In our expressions for $k$, the factors of $1/(4\pi)^2$ arise from three-body phase space, and $g_{abc}$ generically denotes the coupling between the fields $a$, $b$, and $c$, so for instance $g_{\tilde\chi\tilde \ell \ell}$ is the coupling between a neutralino, slepton, and lepton.  The expressions for $g_{abc}$ shown in the caption of Table \ref{tab:FOD1} were computed using a mass mixing insertion approximation.  Moreover, $g_{1a}$ denotes the hypercharge coupling of the field $a$, while $g_{2}$ denotes the $SU(2)$ coupling.  The symbol $\theta_{ab}$ denotes the mixing angle between the fields $a$ and $b$, so for instance $\theta_{\tilde \chi \tilde h}$ is the mixing angle between the neutralino and the pure higgsino.  Because we are concerned with a primarily bino-like neutralino and mostly right-handed slepton, $\theta_{\tilde \chi \tilde w}$, $\theta_{\tilde \chi \tilde h}$, and $\theta_{\tilde \ell \tilde \ell_L}$ are small while $\theta_{\tilde \chi \tilde b}$ and $\theta_{\tilde \ell \tilde \ell_R}$ are order unity.
We have also taken the Higgs vev to be of order $m$ for simplicity in this analysis.  Lastly, note the factors of $m^2/m_{a}^2$ ($m^4/m_{a}^4$) which arise from three-body decays through an off-shell fermionic (bosonic) field $a$.

Let us now discuss the salient features of the tables. To begin, consider the case of a right-handed slepton LOSP, which is relatively straightforward in terms of collider signals.  For the R-parity even operators and the Bino Portal operator, visible sector lepton number is conserved and so the right-handed slepton decays to a charged lepton and an LSP, for example via $\tilde \ell^\pm \rightarrow \ell^\pm \tilde x'$.  If the LOSP is a smuon or selectron, then for the fraction of LOSPs which stop inside the detector before decaying, the outgoing charged lepton is monochromatic and offers the best conditions for reconstructing the LSP mass.

For the remaining R-parity odd operators, visible sector lepton number is violated.  Thus, the slepton LOSP will invariably decay into an even number of visible sector leptons, for instance via $\tilde \ell^\pm \rightarrow \ell^\pm \nu \tilde x'$.  These decays often involve neutrinos, either directly or from decays of charged $W$ and higgs bosons.  Because of the inherent challenge of identifying outgoing neutrinos, this scenario may be more difficult to distinguish from the R-parity even collider signals.  However, in the case of smuon or selecton LOSP it may be possible to distinguish between $\tilde \ell^\pm \rightarrow \ell^\pm \tilde x'$ and $\tilde \ell^\pm \rightarrow \ell^\pm \nu \tilde x'$, based on whether the outgoing charged lepton is monochromatic in events with stopped LOSPs.

Second, let us consider the case of a bino-like neutralino LOSP.  If $X'$ couples via $\mathcal{O}_{K}$ or $\mathcal{O}_{W}$, then there is a two-body decay to a $Z$ boson whose leptonic decays may be useful for event reconstruction and eventual LSP mass measurement.  Moreover, there is a subdominant three-body decay to di-lepton through an off-shell slepton which might be likewise be employed.  Note that it is possible to distinguish $\mathcal{O}_{K}$ from $\mathcal{O}_{W}$ at colliders based on the fact that only $\mathcal{O}_{W}$ allows for a decay to photons via $B^\alpha B_\alpha X'$. 

Next, consider the scenario where $X'$ couples via $B^\alpha X'_\alpha$ or $H_u H_d X'$, corresponding to Bino and Higgs Portals, respectively.  In both cases, the leading decay process of the neutralino occurs via mixing with the LSP.  Specifically, the neutralino two-body decays via its tiny LSP fraction into two hidden sector states, so $\tilde \chi^0 \rightarrow y' \tilde y'$.  However, in both cases there is also a subdominant visible decay mode which may be used to reconstruct the LSP mass.

The invisible neutralino LOSP decays, $\tilde \chi^0 \rightarrow y' \tilde y'$, occur via the Bino and Higgs Portals and can pose a problem for reconstructing FO\&D because more than a single $\tilde x'$ may be produced from the decay of the LOSP.  In general, R-parity only implies that an {\it odd} number of $\tilde x'$ particles are produced.  While $\tilde y'$ ultimately cascades into $\tilde x'$ due to R-parity, $y'\rightarrow \tilde x' \tilde x'$ could yield two additional $\tilde x'$ particles.  The effects of these unwelcome LSP multiplicities on cosmology may be small for two reasons.   First, the coupling controlling the invisible decay may be small.  For example in the case of the Bino Portal, one can demand a smallish gauge coupling for the hidden sector $U(1)$.  This has no effect on the strength of the visible decays of the LOSP, which are fixed by the MSSM hypercharge gauge coupling, but will reduce the branching ratio to invisible decays.  Second, production of multiple LSPs is forbidden by kinematics if $m <3 m'$.

Finally, consider the case where the neutralino LOSP couples to the R-parity odd operators, $LH_u$ and $LH_d^\dagger$.  For these operators the LOSP decays through the higgsino fraction of the neutralino.  However, the leading two-body decay is $\tilde \chi^0 \rightarrow \nu \tilde x'$, which is invisible, and thus useless for LSP mass reconstruction.  The leading visible decay is three-body through an off-shell higgsino-like neutralino, give by $\tilde \chi^0 \rightarrow h^\pm \tilde h^{\mp*} \rightarrow h^\pm \ell^\mp \tilde x'$, and likewise with $h^{\pm}$ replaced with $W^\pm$.   

Lastly, note that modifying the portal operators by inserting powers of supersymmetry breaking spurions leaves the decay modes and $k$ factors of Tables \ref{tab:FOD1} and \ref{tab:FOD2} parametrically unchanged if $m \sim m_{3/2} \sim v$.

\begin{figure}
\begin{center}  \includegraphics[scale=.8]{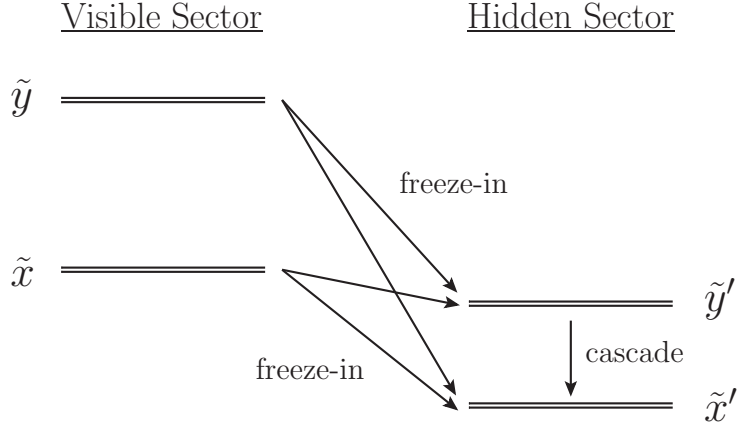}\end{center}
\caption{\footnotesize{Depiction of the cosmological history of FI. While the heavy visible superparters, $\tilde y$, and the LOSP, $\tilde x$, are in thermal equilibrium, they can FI an abundance of the LSP, $\tilde x'$, or a hidden sector spectator, $\tilde y'$.  Due to R-parity each $\tilde y'$ ultimately cascades down to $\tilde x'$.    In order to reconstuct this cosmology we need to be able to measure the lifetime and mass of $\tilde x$ and the mass of $\tilde x'$. }}
\end{figure}

\section{Freeze-In (FI)}
\label{sec:FI}

Let us now consider the FI mechanism whereby $\tilde{x}'$ particles are produced from the decays of visible sector particles \emph{before} they undergo thermal FO.  In this scenario, the dominant contribution to the abundance of $\tilde{x}'$ arises when visible particles become non-relativistic, which is at temperatures of order their mass. 

\subsection{Spectator Fields in Freeze-In}

In FO\&D, particles other than the LOSP and LSP have essentially no effect on the cosmological history of the universe.  In contrast, FI can be substantially affected by spectator fields in the visible and hidden sectors, collectively denoted by $\tilde{y}$ and $\tilde{y}'$. In particular, the decays $\tilde{y} \rightarrow (\tilde{x}',\tilde{y}') + \textrm{SM}$, and 
$\tilde{x} \rightarrow \tilde{y}' + \textrm{SM}$ may occur while these fields are still in thermal equilibrium but after the LSP has frozen out in the hidden sector.  These decays produce additional FI contributions to the final LSP abundance.
The rate of these decays cannot be directly measured at colliders, and generically they cannot be inferred from measurements of the LOSP $\tilde{x}$ decay. Therefore, there is no model-independent way to reconstruct FI at the LHC.

Nevertheless, there can easily be situations in which the FI contributions of $\tilde{y}$ particles to the LSP via $\tilde{y} \rightarrow \tilde{x}' + \textrm{SM}$ may be inferred from the visible decays $\tilde{x}\rightarrow \tilde{x}' + \textrm{SM}$.  On the other hand, the decays $(\tilde{y},\tilde{x}) \rightarrow \tilde{y}' + \textrm{SM}$ are invariably model-dependent and thus unknown, as they require information about the entire hidden sector. In certain cases, however, such decays may be either forbidden (for example, by symmetries), or irrelevant (for example, if decays to $\tilde{y}' + \textrm{SM}$ allowed by symmetries are kinematically inaccessible). This will also be clear from the models discussed in Appendices \ref{higgsportal-model} and \ref{app:bino}. Hence, it will be assumed hereafter that the decays $(\tilde{y},\tilde{x}) \rightarrow \tilde{y}' + \textrm{SM}$ do not provide an important contribution to FI of the LSP.  

From above, it is clear that the criterion for reconstructing hidden sector cosmology from measurements at colliders imposes a constraint on theoretical models realizing the FI mechanism. On the other hand,  within theoretical models satisfying the above constraint, there are very few constraints on the identity of the LOSP---the only requirement being that the LOSP FO abundance provides a subdominant component to the total relic abundance. This allows almost the entire superpartner spectrum of the MSSM to be the LOSP for sub-TeV masses, with the exception of the bino. Note that the situation is complementary to that for FO$\&$D, where the identity of the LOSP is limited to the bino and right-handed sleptons, but there are no or minimal constraints on theoretical models. 

\subsection{Theories of Freeze-In} \label{FI-models}

We now discuss models in which the couplings for all relevant decays for FI such as $\tilde{y}\rightarrow \tilde{x}' + \textrm{SM}$ can be inferred from the measurable decays 
$\tilde{x}\rightarrow \tilde{x}' + \textrm{SM}$ by using the R-symmetry and R-parity transformation properties of the MSSM operators listed in (\ref{eq:ops}), which can couple to $X'$ at dimension $\leq$ 5. 

A useful first step in constructing such models is to consider R-parity and R-symmetry assignments of $X'$ which allow only one operator coupling the visible sector to the hidden sector while forbidding others. One could then hope that all couplings between the visible and hidden sector would proceed via this operator, allowing us to infer the couplings for the decays $\tilde{y} \rightarrow \tilde{x}' + \textrm{SM}$. However, as mentioned earlier, we work within the framework of a low energy effective field theory in which supersymmetry breaking is mediated to us by Planck-suppressed operators with unknown coefficients. Thus, any supersymmetric operator allowed by symmetries can be dressed with functions of supersymmetry breaking spurions $\Phi$ with arbitrary coefficients, which complicates the situation.

If $X'$ is assigned to be R-parity even with R-charge zero, then it can couple via ${\cal O}_{W}X'$ and ${\cal O}_{K}X'$ at a supersymmetric level.  
This is a problem since the different supersymmetric operators have different coefficients in general, only a handful of which may be measured from the decays of the LOSP. Therefore, in order to reconstruct the FI mechanism from measurements at the LHC, it is important to couple an R-parity even chiral superfield $X'$ to a class of operators in the visible sector containing only one operator. There is only one such operator in the list \ref{table:optable} --- $H_u\,H_d$, and giving rise to the connector operator $\lambda\,\int d^2\theta\,H_u\,H_d\,X'$, dubbed the Higgs Portal in Section \ref{sec:opanalysis}. 

One can try to do the same for an R-parity odd chiral superfield $X'$. A simple example is given by the supersymmetric operator $\lambda_i \int d^2\theta\,L\,H_u\,X'$ which could be arranged to be the only supersymmetric operator allowed by R-parity and R-symmetry. Since this operator is linear in lepton fields, its operator coefficient, $\lambda_i$ has a lepton flavor index.  Thus, for a slepton LOSP, the prospect of reconstructing all of the operator coefficients is not promising.  For example, if the LOSP is a stau, then  $\lambda_3$ may be measured, but $\lambda_{1,2}$ will be inaccessible at colliders.  On the other hand, for higgsino LOSP, the above operator gives rise to $\tilde{h}_u\rightarrow \ell\,\tilde{x}'$.  By measuring the branching ratios of decays into each lepton generation, $\lambda_i$ can be fully measured.  
  
Nevertheless, this theory still has a problem since there exist supersymmetry breaking operators derived from the supersymmetric operator which give rise to additional unknown contributions to FI in the early Universe. For example, if the symmetries allow $\int d^2\theta LH_u X'$, then they also allow the supersymmetry breaking operator $\int d^4\theta\,L\,H_u\,X'\frac{\Phi^{\dag}\Phi}{M_{pl}^2}$.  This operator mediates the process $\tilde{\ell}\rightarrow h\,\tilde{x}'$.  Thus, in this scenario there are at least two active FI processes, each involving a different superpartner, $\tilde h$ and $\tilde \ell$ with different unknown decay widths.  Since there is only one LOSP, only one of these decay processes can be measured at the LHC, making reconstruction of the full FI mechanism difficult.  Note that this is not an issue for R-parity even $X'$ which couples to the visible sector, such as via $H_u\,H_d\,X'$ above,  since then $\tilde{x}'$ is a fermion and the induced supersymmetry breaking operator does not give rise to a decay to $\tilde{x}'$ from an R-parity odd particle in the visible sector. 

Until now, the LSP was assumed to be a chiral superfield. If the LSP is a $U(1)'$ vector superfield $X_{\alpha}'$, then it can only couple via the operator $B^\alpha X'_\alpha$, consisting of the single operator $B_{\alpha}$ and giving rise to the Bino Portal operator $\lambda\,\int  d^2\theta\,B^{\alpha}\,X_{\alpha}'$.   This kinetic mixing can be removed by a shift of the gauge fields, inducing a $U(1)'$ millicharge for MSSM fields.  Hence, this single kinetic mixing operator uniquely fixes the couplings of {\it every} MSSM superpartner to $\tilde x'$.  As such, from the measurement of the decay of the LOSP, the coupling of $\tilde x'$ to the remaining MSSM particles can be inferred.  With these couplings, the contribution to FI from all the MSSM fields can be computed, and checked against the FI formula in Eq.~\ref{eq:FIeq}.  Thus, the FI mechanism can be fully reconstructed from measurements at the LHC for these models. 

\subsection{UV Sensitivity of Freeze-In} \label{FI-UV}

If DM is dominantly produced by FI, its abundance will only be reconstructable if the FI process is IR dominated. 
FI of $X'$ can occur via both decays and scatterings of visible sector particles.  While the decay contribution is always IR dominated, the contribution from scattering at temperature $T$ to the FI yield has the behavior $Y_d(T) \propto T^{2d-9}$, where $d$ is the dimension of the portal operator inducing FI.  Thus, for the $d=4$ operators of Table \ref{table:optable} all contributions to FI are IR dominated, and this includes both Bino and Higgs Portals.   On the other hand, for $d \geq 5$, FI by scattering is UV dominated.  However, any production of $X'$ before $X'$ freeze-out is irrelevant to the final yield of DM hence, for the $d=5$ operators of Table \ref{table:optable}, the question is whether FI by decay beats FI by scattering after $X'$ freeze-out.

For the $d=5$ operators $(\mathcal{O}_K, \mathcal{O}_W, L H_d^\dagger, L^\dagger H_d, L^\dagger H_u^\dagger )X'$ there are contributions to FI from 2-body decay modes, and these dominate the scattering contribution provided the temperature of the visible sector when $X'$ freezes out is less than about 10 TeV \cite{Hall:2009bx}, which is always the case \cite{cosmopaper}.
On the other hand, for the operators $(LLE, QLD, UDD)X'$ the decays are always at least three body, giving phase space suppression relative to scattering, so that FI by scattering can not be ignored even if $X'$ FO$'$ occurs as late as $T \simeq m$.   Hence, the FI abundance resulting from these operators is always UV sensitive.

\begin{table}[t!]
\centering
\small\addtolength{\tabcolsep}{-5pt}
\renewcommand{\arraystretch}{1.8}
\begin{center}
\begin{tabular}{||c||c|c||c|c||}
\hline\hline
\multicolumn{5}{|c|}{FI} \\
\hline
\hline
& \multicolumn{2}{|c||}{Higgs Portal: $ H_u H_d X^\prime $} 
& \multicolumn{2}{|c||}{Bino Portal: $B^\alpha X_{\alpha}^\prime $} \\ 
\hline
LOSP & Decay & k & Decay & k\\
\hline
$\tilde{g}$ & $\tilde{g} \rightarrow q q \tilde{x}^{\prime}$ & $\frac{1}{(4 \pi)^2}g_{\tilde h \tilde q q}^2 \frac{m^4}{m_{\tilde q}^4}$ 
& $\tilde{g} \rightarrow q q \tilde{x}^{\prime}$ & $\frac{1}{(4 \pi)^2} g_{1q}^2 \frac{m^4}{m_{\tilde q}^4}$ \\
\hline
\multirow{2}{*}{$\tilde{\nu}$ } 
& $\tilde{\nu} \rightarrow \ell^{\pm}(h^{\mp},W^{\mp})\tilde{x}^{\prime}$ 
& $\frac{1}{(4 \pi)^2}g_{\tilde h \tilde \nu \ell}^2 
\frac{m^2}{m_{\tilde{h}}^2} (1,g_{2}^2)$ 
& $\tilde{\nu} \rightarrow \ell^{\pm} (h^{\mp},W^\mp) \tilde{x}^{\prime}$ 
& $\frac{1}{(4 \pi)^2}g_{1h}^2g_{\tilde h \tilde \nu \ell}^2 
\frac{m^2}{m_{\tilde{h}}^2} (1,g_{2}^2)$   \\ 
& $\tilde\nu \rightarrow \tilde\nu \tilde{x}^{\prime}$ & $g_{\tilde h \tilde \nu\nu}^2$ 
& $\nu \rightarrow \nu \tilde{x}^{\prime}$ & $g_{1 \nu}^2$  \\
\hline
\multirow{2}{*}{$\tilde{q}$} 
& $\tilde{q} \rightarrow  q \tilde{x}^{\prime}$ 
& $g_{\tilde h \tilde q q}^2$ 
& $\tilde{q} \rightarrow q \tilde{x}^{\prime}$ & $g_{1 q}^2$ \\ 
& $\tilde{q} \rightarrow q (h^{0,\pm},W^{0,\pm}) \tilde{x}^{\prime}$ 
& $\frac{1}{(4 \pi)^2}g_{\tilde h \tilde q q}^2\frac{m^2}{m_{\tilde{h}}^2} (1,g_{2}^2)$ 
& $\tilde{q} \rightarrow q (h^{0,\pm},W^{0,\pm}) \tilde{x}^{\prime}$  
& $\frac{1}{(4 \pi)^2} g_{1h}^2 g_{\tilde h \tilde q q}^2\frac{m^2}{m_{\tilde{h}}^2} (1,g_{2}^2)$ \\
\hline
\multirow{2}{*}{$\tilde{\chi}^{\pm}$} 
& $\tilde{\chi}^{\pm} \rightarrow (h^{\pm},W^{\pm}) \tilde{x}^{\prime}$ 
& $g_{2}^2(\theta_{\tilde \chi \tilde w}^2, \theta_{\tilde \chi \tilde h}^2)$
& $\tilde{\chi}^{\pm} \rightarrow (h^{\pm},W^{\pm}) \tilde{x}^{\prime}$
& $g_{1h}^2(\theta_{\tilde \chi \tilde h}^2, \theta_{\tilde \chi \tilde w}^2)$
\\ 
& $\tilde{\chi}^{\pm} \rightarrow \ell^{\pm} \nu \tilde{x}^{\prime}$
& $\frac{1}{(4 \pi)^2} g_{\tilde \chi \tilde \ell \nu}^2 g_{\tilde h \tilde \ell \ell}^2 \frac{m^4}{m_{\tilde{l}}^4}$
& $\tilde{\chi}^{\pm} \rightarrow \ell^{\pm} \nu \tilde{x}^{\prime}$
& $\frac{1}{(4 \pi)^2} g_{\tilde \chi \tilde \ell \nu}^2 g_{1 \ell}^2 \frac{m^4}{m_{\tilde{l}}^4}$
\\
\hline
\multirow{3}{*}{$\tilde\chi_0$} 
& $\tilde\chi_{0} \rightarrow (h^0,Z)\tilde{x}^{\prime}$
& $\theta_{\tilde\chi \tilde h}^2, \theta_{\tilde\chi \tilde h}^2 g_{2}^2$ 
& $\tilde\chi_{0} \rightarrow (h^0,Z)\tilde{x}^{\prime}$
& $\theta_{\tilde\chi \tilde h}^2\,g_{1h}^2, \theta_{\tilde\chi \tilde h}^2 g_{2}^2 g_{1h}^2$ \\ 

& $\tilde\chi_0 \rightarrow y^{\prime} \tilde{y}^{\prime}$ 
& $\theta_{\tilde{\chi}\tilde{h}}^2 \lambda^{\prime 2}$ 
& $\tilde\chi_0 \rightarrow y^{\prime} \tilde{y}^{\prime}$  
&  $\theta_{\tilde{\chi}\tilde{b}}^2 g^{\prime 2}$\\
& $\tilde\chi_{0} \rightarrow \ell^+ \ell^- \tilde{x}^{\prime}$ 
& $\frac{1}{(4 \pi)^2} g_{\tilde\chi \tilde\ell\ell}^2 g_{\tilde{h}\tilde\ell \ell}^2 \frac{m^4}{m_{\tilde{l}}^4} $ 
& $\tilde\chi_{0} \rightarrow \ell^+ \ell^- \tilde{x}^{\prime}$  
& $\frac{1}{(4 \pi)^2} g_{\chi \tilde\ell\ell}^2 g_{1 \ell}^2 \frac{m^4}{m_{\tilde{l}}^4} $ \\
\hline
$\tilde{\ell}^\pm$
& $\tilde{\ell}^\pm \rightarrow \ell^\pm \tilde{x}^{\prime}$ 
& $g_{\tilde h \tilde\ell \ell}^2$ 
& $\tilde{\ell}^\pm \rightarrow \ell^\pm \tilde{x}^{\prime}$ 
& $g_{1\ell}^2$ \\ \cline{2-5}
\hline\hline

\end{tabular}
\end{center}
\caption{\footnotesize{Summary of LOSP decay modes and rates for FI for interactions corresponding to the Higgs and Bino Portals.   Here $g_{\tilde h \tilde q q}\equiv \lambda_{q} + v g_{2}^2/m_{\tilde w}+ v g_{1q}g_{1h}/m_{\tilde b}$ is the effective coupling between the higgsino $\tilde h$ and $\tilde q q$ and $g_{\tilde h \tilde \nu \nu}=v g_{2}^2/m_{\tilde w}+ v g_{1l}g_{1h}/m_{\tilde b}$ is the effective coupling between higgsino and $\tilde{\nu} \nu$. $g_{\tilde{\chi}\tilde{\ell}\nu}$ and $g_{\tilde{h}\tilde{\ell}\nu}$ are defined similarly as $g_{\tilde{\chi}\tilde{\ell}\ell}$ and $g_{\tilde{h}\tilde{\ell}\ell}$ in the caption of Table \ref{tab:FOD1}. Also, $y'$ and $\tilde y'$ denote hidden sector states.  The dimensionless branching ratio, $k$, is defined in Eq.~\ref{eq:kdef}}}\label{tab:FI}
\end{table}

\subsection{Collider Signatures of Freeze-In}\label{FI-signals}

Table \ref{tab:FI} lists the important decay modes for all possible LOSP candidates for the Bino and Higgs Portal models, in which measurements at the LHC can reconstruct the FI mechanism in the early Universe. As mentioned earlier, a variety of LOSP candidates are possible---gluinos, squarks, sleptons, charginos, neutralinos and sneutrinos. Most of them have more than one interesting decay mode, as can be seen from Table \ref{tab:FI}. The table shows all two-body decay modes.  Lepton rich three-body modes are also shown, as they are more useful than hadronic modes for determining the LSP mass at the LHC. Three-body hadronic modes and those with Higgs and vector bosons are only shown if two body modes are absent or are hard to measure.  Table \ref{tab:FI} uses the same notation as in Tables \ref{tab:FOD1} and \ref{tab:FOD2}, which is discussed in Section \ref{FOD-signals}.

Some salient features of Table \ref{tab:FI} are worth mentioning. First, it is striking that in both the Higgs and Bino portal, the decay modes for all LOSP candidates are exactly the same, albeit with different branching ratios for the various modes in general.   This is the case because the basic effect of both the Higgs and Bino portal is to induce mass/kinetic mixings among the neutralinos and $\tilde x'$.   This will be important in distinguishing the models. Note that many channels can proceed through both gauge couplings ($g_1,g_2$) and Yukawa couplings ($\lambda_q,\lambda_l$); hence many of the branching ratios depend on both of them. The slepton LOSP has a two body leptonic mode, so it should be quite easy to measure. Chargino LOSPs have a two-body mode containing $h^{\pm}$ or $W^{\pm}$, and the LSP mass could probably be measured through the leptonic decays of $h^{\pm},W^{\pm}$. The leading decays of squark LOSPs are hadronic, but the LSP mass should be measurable since the decay is two-body and the jet is expected to be hard. 

For neutralino LOSPs, apart from the two-body decays with $h,Z$ and three-body decays with $l^{+}l^{-}$ which should be measurable, there is also a completely invisible mode with both decay products in the hidden sector. This is made possible due to the mixing between $\tilde{\chi}^0$ and $\tilde{x}'$ for both the Bino and Higgs Portals. However, as long as the visible decay width is sufficiently large so as to be measurable, it should be possible to measure the mass of the LSP. The sneutrino LOSP also has a practically invisible two-body decay mode to $\nu\,\tilde{x}'$; therefore one has to rely on the three-body modes to measure the mass of the LSP in this case.

The completely invisible decay mode for a neutralino LOSP gives rise to some subtleties for the FI mechanism, similar in spirit to  those for LOSP FO$\&$D discussed in Section \ref{sec:FOD}.  The FI abundance from the LOSP depends on the total LOSP decay width, but the LOSP lifetime is typically too long to be measured from the distribution of decay lengths in events at the LHC. Hence, the invisible decay mode induces an \emph{a priori} unknown contribution to the FI abundance since the decay width for this mode depends on hidden sector couplings collectively denoted by $g'$. In addition, the invisible decays could give rise to more than one LSP $\tilde{x}'$, in general an odd number of $\tilde{x}'$ due to R-parity. This could occur via $\tilde{\chi}^0 \rightarrow y'\,\tilde{y}'$, with $\tilde{y}'$ ultimately decaying to one $\tilde{x}'$ and $y'$ decaying to two $\tilde{x}'$ if kinematically allowed. 

However, these invisible decays will be subdominant as long as the hidden sector couplings are not too large, for example if $g'$ is comparable to the visible sector gauge couplings. This is because the FI  contributions from the other decays of $\tilde{\chi}^0$ as well as from all the decays of other superpartners only depend on the parameter $\lambda$ (and visible masses and mixings) which can be measured from the visible decays of $\tilde{\chi}^0$. Thus, the unknown contribution is one among many known contributions. Assuming for simplicity that $g' \simeq g_{1,2}$, the branching ratio for the invisible decays is given roughly by:
\begin{equation} 
B_{\rm invis} \sim \frac{1}{N_{\rm eff}}
\label{eq:Binvis}
\end{equation} 
where $N_{\rm eff}$ varies from unity, when the non-LOSP visible superpartners are very heavy and their contributions to FI are negligible, to $N_{\rm eff} \sim 60$ when these non-LOSPs are close in mass to the LOSP and each superpartner gives a comparable contribution to FI.\footnote{In passing, we note that Figure \ref{fig:introplot} is drawn for $N_{\rm eff}=1$.}   Hence, although spectrum dependent, the single unknown contribution to FI from $\tilde \chi_0 \rightarrow y' \tilde y'$ is typically negligible.  

Is it possible to distinguish between the two models for LSP FI? The answer is yes.  The reason is that in the Higgs and Bino Portals, $\tilde x'$ is effectively an additional neutralino which is very weakly coupled to the MSSM.  Because the Higgs (Bino) Portal mediates mass/kinetic mixings between $\tilde x'$ and the higgsino (bino), the $\tilde x'$ is higgsino (bino)-like.  This implies an important difference in the branching ratios to various decay modes of the LOSP.  For example, consider the example of a chargino LOSP.  As shown in Table \ref{tab:FI}, if the chargino is wino-like, then for the Higgs (Bino) Portal the dominant decay mode is to $h^\pm$ ($W^\pm)$.  On the other hand, if the chargino is higgsino-like, then for the Higgs (Bino) Portal the dominant decay mode is to $W^\pm$ ($h^\pm$).  Hence, the branching ratios of the LOSP can be used to infer whether the connector interaction is the Higgs or Bino Portal.  More generally, for the decay branching ratios for squark, slepton, neutralino, and gluino LOSP can also be used to determine the connector operator.

Finally, it is worth noting that a gluino LOSP gives a signal very similar to that for split-supersymmetry \cite{ArkaniHamed:2004yi,ArkaniHamed:2004fb}, namely a three-body decay mode, $\tilde x \rightarrow q\bar{q}\tilde{x}'$. However, it should be easy to distinguish the FI scenario from split-supersymmetry. In split-supersymmetry the LOSP is generically not the gluino, and furthermore the squarks are too heavy to be produced at the LHC. In the FI scenario on the other hand, it is possible to produce squarks and have a long-lived gluino decay signal. 

\section{Detection Prospects at the LHC}\label{detect}

What are the prospects for detection of these particles at the LHC? This depends crucially on the nature of the LOSP---in particular, whether it is neutral or charged/colored. 
There is a large body of literature on  detecting long-lived charged particles (such as staus) \cite{CMS-stopped,Hamaguchi:2004df} and colored particles (such as gluinos) \cite{Arvanitaki:2005nq}, but relatively very little on long-lived neutral particles, however see \cite{Meade:2010ji}. This is because charged and colored long-lived particles (LOSPs in our case) leave tracks in the detector and can be stopped effectively by their interactions with matter. Experimental searches for neutral LOSPs, on the other hand, are quite difficult and depend strongly on the lifetime of the LOSP as the lifetime determines the number of LOSPs decaying inside the detector. 

\subsection{Charged or Colored LOSP}\label{charged}

Charged and colored LOSPs have lifetimes and decay products that can be measured for a huge range of lifetimes, from $10^{-12}$ s to at least $10^6$ s. Furthermore, charged sleptons and charginos will produce spectacular and distinctive massive charged tracks at the LHC allowing a measurement of the LOSP mass \cite{Meade:2010ji}. Some fraction of LOSPs, emitted with small velocities, will lose their kinetic energy by ionization and stop inside the calorimeter. There have been proposals by both ATLAS and CMS to use the existing detector apparatus to precisely measure LOSP decays in their ``inactive" mode \cite{CMS-stopped, Asai:2009ka, Arvanitaki:2005nq}. In addition, there have been numerous proposals for stopper detectors which may be built outside the main detector to stop these particles in order to perform precision spectroscopy on their decays \cite{Hamaguchi:2004df}. These stopper detectors could perform extremely precise measurements of the lifetime of the LOSP as well as the mass of the LSP, assuming that it is sufficiently heavy, $m' > 0.2\,m$  for a stau LOSP. A similar analysis applies for squark and gluino LOSPs, relevant for FI. 

For the FI scenario, it is important to measure the coupling $\lambda$ for the Higgs and Bino Portals, discussed in Section \ref{sec:FI}. In the limit where the hidden sector coupling $g'$ is small, $\lambda$ can be extracted by measuring the total lifetime of the LOSP.
If $g'$ is not small, so that the invisible branching ratio is relevant to the extraction of $\lambda$, one can make progress by the following procedure. R-parity implies that all supersymmetric events end up with two LOSPs.  One can compare the number of events with one invisible decay and one visible decay of the LOSP with the number of events with both LOSPs decaying visibly. This gives the ratio of the invisible and visible decay widths and, combining with the previous procedure, allows a measurement  of $\lambda$. 

\subsection{Neutral LOSP}\label{neutral}

As mentioned earlier, the prospects for neutral LOSPs depend crucially on their lifetime, which has a different range in the FO$\&$D and FI scenarios.  Since the FI mechanism gives rise to a relic abundance proportional to the decay width of LOSP (and the partial width of other superpartners to the LSP), requiring that FI gives the total relic abundance of the LSP essentially fixes the lifetime of the LOSP ($\tilde{\chi}^0$ or $\tilde{\nu}$) to be $\sim 10^{-2}$\,s, giving a decay length $L\equiv \gamma c\tau$ of the LOSP of
 \begin{equation}
 L_{\rm FI} \sim 10^6 \textrm{ meters} \times \gamma \; \left(\frac{m'/m}{0.25}\right) \left(\frac{300 \mbox{GeV}}{m}\right) \;  \frac{1}{N_{\rm eff}}\;  
\end{equation}
where $N_{\rm eff} >1$ arises from the FI contribution of non-LOSPs, as described by Eq. (\ref{eq:Binvis}).

On the other hand, for FO$\&$D the lifetime of the LOSP (bino-like $\tilde{\chi}^0$) is not relevant for the relic abundance, only its mass. As shown in Figure \ref{fig:introplot} the lifetime for FO\&D varies very widely from less than about 100 s from nucleosynthesis to greater than about  $\sim 10^{-13}$ s from the requirement that the two sectors \emph{not} be in thermal equilibrium with each other, giving
\begin{equation}
L_{\rm FO\&D} \sim (10^{10} - 10^{-5} \textrm{ meters}) \gamma.
\end{equation}
Note that for $10^{-8}\,{\textrm{ s}} < \tau < 10^{-2}\textrm{ s}$, we are in the region where the contribution from FI is above the critical abundance giving rise to reannihilation of LSPs in the hidden sector and that the hidden sector annihilation cross-section is large enough so that LOSP FO$\&$D provides the dominant relic abundance; see Section \ref{sec:FOD} for details. 

Given the number of LOSPs produced at the LHC ($N_{\rm produced}$), the number of LOSPs decaying within the detector ($N_{\rm decayed}$) is given by:
\begin{equation}\label{ndecay}
N_{\rm decayed} = N_{\rm produced}\,(1-e^{-d/L}) \stackrel{L \gg d}{\rightarrow}  N_{\rm produced}\,\frac{d}{L}
\end{equation}
where $d$ is the size of the detector ($\sim 10$ m). Assuming a typical strong production cross-section of ${\cal O}(100)$ pb at the LHC gives $N_{\rm produced}\approx 10^7$ with an integrated luminosity of 100 fb$^{-1}$. Thus, from (\ref{ndecay}) a signal of $N_{\rm decayed}\gtrsim 10^2$ requires:  
\begin{equation}
L \lesssim 10^6\;\mathrm{meters}.
\end{equation}
Thus, we find that a large fraction of the lifetime range for FO\&D from the decay of a bino-like $\tilde{\chi}^0$ can be probed at the LHC.  However, the situation for FI with a neutral LOSP, $\tilde{\chi}^0$ or $\tilde{\nu}$, is critical, and is likely to be seen only for high luminosities and high values of $m$, $m/m'$, and $N_{eff}$.

\section{Discussion}\label{sec:con}

Early running at the LHC will test a wide class of weak-scale supersymmetric theories that have unconventional DM: an LSP residing in a hidden sector.   Any such two-sector cosmology has a long-lived LOSP, with a lifetime greater than $\tau_{\rm min} \simeq 10^{-13}$ s so that, for a total superpartner production cross section of 1-10 pb at the LHC with $\sqrt{s}$ = 7 TeV, an integrated luminosity of 1 $\mbox{fb}^{-1}$ will yield $10^{3-4}$ events with long-lived LOSPs.  If such events are observed, how can we deduce the underlying cosmology? 

The visible and hidden sectors have different temperatures, allowing two classes of thermal relic abundance mechanisms:  conventional Freeze-Out of the LOSP followed by LOSP decay, and Freeze-In of the LSP from visible superpartners as they become non-relativistic.   In this paper we have concentrated on the case that these processes are symmetric between particles and anti-particles, and that the production of the LSP is not so copious as to initiate re-annihilations of LSPs in the hidden sector, giving the cases of FO\&D and FI.  In these cases the DM abundances are tightly related to measurable parameters, as shown in Figure \ref{fig:introplot}, and we summarize our results in the next sub-section.  However, in two sector cosmologies both FO\&D and FI come in versions with particle anti-particle asymmetries and with re-annihilations.  These  cases  \{FO\&D$_{\rm r}$, FO\&D$_{\rm a}$, FI$_{\rm r}$, FI$_{\rm a}$\} occur for a wide range of parameters and, although a complete reconstruction of the DM density is difficult, the collider signals are distinctive and are summarized in the second sub-section.

\begin{table}
\centering
\small\addtolength{\tabcolsep}{-5pt}
\renewcommand{\arraystretch}{1.8}
\begin{center}

\begin{tabular}{||c||c|c|c|c||}
\hline
\hline

\multirow{3}{*}{} & $L^\dagger L X^\prime$ & & $QUHX^\prime$ & \\
& $Q^\dagger QX^\prime$ & $W^2 X^\prime$ & $LEHX^\prime$ & $H_u H_d X^\prime$  \\
& $H^\dagger HX^\prime$ & & $QDHX^\prime$ & \\
\hline
\hline
$\tilde{\chi}^0$ & $h^0 ,Z ,\ell^+ \ell^-$ & $\gamma$, $Z$ & $l^+ l^-$ & $h^0,Z,l^+l^-$ \\
\hline 
$\tilde{l}^\pm$ & $l^\pm$ & $l^\pm \left(\gamma,Z,h^0\right),\nu( W^\pm,h^\pm) $ & $l^\pm$ & $l^\pm$ \\
\hline
\hline
$\tilde{\chi}^\pm$ & $h^{\pm}, W^\pm, \ell^\pm \nu$ & $h^\pm,W^\pm$ & $l^\pm \nu$ & $h^\pm, W^\pm$  \\
\hline
$\tilde{\nu}$ & $\nu (1, h^0 , Z), l^\pm (h^\mp W^\mp) $ & $\nu\left(\gamma,Z,h^0\right), \ell^\pm (W^\mp, h^\mp)$ & $l^\pm (h^\mp,W^\mp)$ 
& $\nu (1, h^0 , Z), l^\pm (h^\mp W^\mp)$  \\
\hline
$\tilde{q}$ & $j(1,h^0,Z, h^\pm ,W^\pm)$ & $j \left( \gamma, Z, h^0, W^\pm, h^\pm \right)$ & $j(1,h^0,Z, h^\pm ,W^\pm)$ & $j(1,h^0,Z, h^\pm ,W^\pm)$ \\
\hline 
$\tilde{g}$ & $jj (1,h^0,Z, h^\pm ,W^\pm)$ & $jj \left( \gamma, Z, h^0, W^\pm, h^\pm \right)$ & $jj(1,h^0,Z, h^\pm ,W^\pm)$ & $jj(1,h^0,Z, h^\pm ,W^\pm)$  \\
\hline 
\hline

\end{tabular}
\end{center}
\caption{\footnotesize{Signal topologies at displaced vertices from LOSP decays induced by R-parity even $X'$, for a variety of LOSP candidates.    A jet is represented by $j$.  All topologies have missing energy carried by the LSP.}}
\label{tab:REsig}
\end{table}

\begin{table}
\centering
\small\addtolength{\tabcolsep}{-5pt}
\renewcommand{\arraystretch}{1.8}
\begin{center}

{\footnotesize
\begin{tabular}{||c||c|c|c|c|c|c||}
\hline
\hline

\multirow{3}{*}{} & &  $LH_u X^\prime$ & $L H_d^{\dagger} X^\prime$ & & &\\
& $B^\alpha X_\alpha^\prime$ & & & $LLEX^\prime$ & $QDLX^\prime$ & $UDDX^\prime$ \\
& & $L H_u X^{\prime\dagger}$ & $L H_d^\dagger X^{\prime\dagger}$ & & &\\
\hline
\hline
$\tilde{\chi}^0$ 
& $h^0,Z,l^+ l^-$ & $\nu(1,h^0,Z), l^\pm(h^\mp,W^\mp) $ & $\nu(1,h^0,Z), l^\pm(h^\mp,W^\mp) $
& $l^+ l^- \nu$ & $j j \left(l^\pm, \nu \right)$ & $jjj$ \\
\hline 
$\tilde{l}^\pm$ & $l^\pm$ & $h^\pm,W^\pm$ & $h^\pm,W^\pm$ 
& $l^\pm \nu$ & $jj$ & $jjj\left( l^\pm, \nu \right)$\\
\hline
\hline
$\tilde{\chi}^\pm$ & $h^\pm W^\pm$ & $l^\pm$ & $l^\pm$ & $l^\pm l^+ l^-$, $l^\pm \nu\nu$ 
& $jj \left(l^\pm,\nu \right)$ & $jjj$ \\
\hline
$\tilde{\nu}$ & $\nu (1, h^0 , Z), l^\pm (h^\mp W^\mp)$ & $h^0,Z$ & $h^0,Z$ & $l^+ l^-$ & $jj$ 
& $jjj \left(l^\pm , \nu \right)$ \\
\hline
$\tilde{q}$ & $j(1,h^0,Z, h^\pm ,W^\pm)$ & $j(l^\pm,\nu)$& $j(l^\pm,\nu)$ 
& $j\left(l^+l^-\nu, l^\pm l^+l^-,l^\pm \nu\nu \right) $ & $j \left(l^\pm,\nu\right)$ & $jj$\\
\hline 
$\tilde{g}$ &  $jj(1,h^0,Z, h^\pm ,W^\pm)$ & $jj (l^\pm,\nu)$ 
& $jj (l^\pm,\nu)$ &$jj\left(l^+l^-\nu, l^\pm l^+l^-,l^\pm \nu\nu \right) $ & $jj\left(l^\pm,\nu\right)$ & $jjj$ \\
\hline 
\hline

\end{tabular}
}

\end{center}
\caption{\footnotesize{Signal topologies at displaced vertices from LOSP decays induced by R-parity odd $X'$, for a variety of LOSP candidates.  A jet is represented by $j$.  All topologies have missing energy carried by the LSP.}}
\label{tab:ROsig}
\end{table}


\subsection{Signals from Freeze-Out and Decay and Freeze-In}
If superpartners are produced at the LHC, standard analyses of cascade chains will allow the determination of both the nature and mass of the LOSP.  If the LOSP is a neutralino or charged slepton it will be important to make sufficient measurements to infer the annihilation cross section, $\langle \sigma v\rangle$, to determine whether they are candidates to yield DM by FI or FO\&D.  Other LOSP candidates accessible to LHC will be candidates for yielding DM by FI.

What operator is relevant for inducing the LOSP decay?  In Tables \ref{tab:REsig} and \ref{tab:ROsig} we give the topologies for the displaced decay vertices that arise for all LOSP candidates and all decay operators of dimension 4 and 5 when the visible sector is the MSSM.  Once the nature of the LOSP is known, one can examine the corresponding row of the tables.  Sometimes this gives a unique result:  for example a chargino LOSP with a dominant decay to $l^+ l^- l^\pm \tilde{x}'$ implies DM by FI via the operator $LLEX'$; a charged slepton LOSP with dominant decay to two jets plus missing energy is decaying via the $QDLX'$ operator.  Depending on the properties of the slepton, DM may arise from FO\&D or FI.  On the other hand, in many cases the decay topology does not select a unique operator, but leaves ambiguities.  For example a light charged slepton LOSP, with a two body decay to a charged lepton and the LSP, would be a signal of  DM by FI, but could occur via either the Higgs or Bino Portal, or indeed from the $L^\dagger L X'$ or $LEH X'$ operators.  In the next subsection we imagine a particular set of measurements at LHC and discuss how this ambiguity could be resolved.  A similar analysis would be necessary for any signal of the Higgs and Bino Portals since, no matter what the LOSP, the same decay signatures occur.   These portals are distinguished by whether the LSP is dominantly mixed with a Higgsino or bino. 

For FO\&D, once the LOSP mass and annihilation cross section are known, reconstruction depends on the ability to measure the LSP mass, and this depends on the decay mode, as discussed in section \ref{sec:FOD}.  The key is to verify the prediction given in the last row of Table \ref{tab:introtab}.   Although not needed for reconstruction, it may be possible to infer the size of the coefficient of the operator inducing the LOSP decay, $\lambda$ of Eq. \ref{eq:lambdadef}.   All possible LOSP decay topologies for FO\&D are shown in Tables \ref{tab:FOD1} and \ref{tab:FOD2}, together with the mixing angles and kinematical factors, $k$, that enter the computation of the LOSP decay rate.  The quantities in $k$ are all visible sector quantities, and measuring them allows $\lambda$ to be deduced from Eq. \ref{eq:kdef} once the LOSP lifetime is measured.

For reconstructing FI it is critical to measure the LSP mass, as discussed in Section \ref{sec:FI}, as well as the LOSP lifetime.  If the dominant contribution to the FI abundance arises from the decay of the LOSP, as might happen if the LOSP is significantly lighter than other superpartners, then the key relation to test is given in the last row of Table \ref{tab:introtab}.  However, the FI contribution from heavier visible superpartners might dominate, and it is in this case that the Higgs and Bino Portals are important.  To compute the FI contribution from the heavier superpartners one must deduce the coefficient $\lambda$ of the portal operator, which can be done with the use of Table  \ref{tab:FI}.  For the LOSP decay mode under study one must measure the corresponding quantity $k$, then a measurement of the LOSP lifetime allows $\lambda$ to be deduced from Eq. \ref{eq:kdef}.  Knowing the portal operator and its coefficient one can then compute the FI abundances from all non-LOSP superpartners and reconstruct the cosmology.

\subsection{An example:  $\tilde{l}^\pm \rightarrow l^\pm \tilde{x}'$  }
To illustrate our ideas, and the challenges involved in reconstructing the DM cosmology, we consider a very specific example.  Suppose that LHC discovers a charged slepton LOSP with mass of 200 GeV.  Frequently the slepton will get stopped in the detector and we suppose that it is measured to have a lifetime of 0.1 s and that it has a dominant two body decay to $l^\pm \tilde{x}'$; further we suppose that $m'$ is reconstructed from these two body decays to be 100 GeV.  How would we interpret this within a two sector cosmology?

First, the low mass of the slepton implies that its FO yields a low abundance; one that would not give sufficient DM if it were stable, so that the mechanism of interest is FI rather than FO\&D.   Second, a glance at the second row of Tables  \ref{tab:REsig} and \ref{tab:ROsig} shows that this signal could result from both Higgs and Bino Portals, and also from the operators  $L^\dagger L X'$ and $LEH X'$. Thirdly, if the slepton LOSP decay was solely responsible for DM FI then, from the relation in the last line of Table \ref{tab:introtab}, it should have a lifetime of $10^{-2}$ s.  Since the measured lifetime is an order of magnitude longer, only 10\% of the observed DM abundance arises from FI from LOSP decays.  Such a direct experimental verification of the process that generated the DM in the early universe would be an exciting triumph, and using LHC data to infer the DM abundance within an order of magnitude would suggest that FI is the right mechanism.   But could further measurements reveal that the remaining 90\% of DM production arose from FI via decays of non-LOSPs?

Consider first the Higgs and Bino Portals, where the symmetries of $X'$ imply that there is a single portal operator with a single coupling $\lambda$.  In these portals, the LOSP decays via a small mass mixing between the DM and the Higgsino or bino that is proportional to $\lambda$.  The crucial coupling $\lambda$ can be extracted from the slepton lifetime by measuring sufficient parameters of the visible neutralino mass matrix.  Knowing $\lambda$ one can then compute the yield of DM from FI from decays of other superpartners, such as the neutralinos, charginos and squarks.  These yields will differ in the Higgs and Bino Portals, and the question is whether sufficient measurements of the visible superpartner spectrum can be made to show that in one of these portals the yield from non-LOSP decays accounts for the remaining 90\% of DM.  If so this would be a remarkable achievement.  

It may be that the superpartner spectrum will be sufficiently determined to exclude the Higgs and Bino Portals.  In this case it would seem that the observed slepton decays are being induced by $L^\dagger L X'$ or $LEH X'$.  It could be that these operators happen to have larger coefficients, $\lambda$, than other operators with the same $R$-parity and $R$ change, so that inferring $\lambda$ from the slepton lifetime allows a successful computation of the full DM abundance from non-LOSP decays.  However, it is likely that the total FI abundance from this operator is insufficient, and that other operators with the same $R$-parity and $R$ charge are making important contributions that cannot be reconstructed.  For example, the $QUH X'$ operator could yield a substantial FI contribution from squark decays; but the corresponding squark branching ratio would be too small to measure.  Such operators would also lead to subdominant decay modes of the LOSP, such as $\tilde{l}^\pm \rightarrow l^\pm \bar{q} q \tilde{x}'$, but the branching ratio may be too small to measure.

Until now we have not discussed the flavor structure of the LOSP and its decays.  To the extent that the charged lepton Yukawa coupling are small, the Higgs and Bino Portals will have $\tilde{l}^\pm_i \rightarrow l^\pm_i \tilde{x}'$; the flavor of the lepton appearing at the LOSP vertex will be the same as the flavor of the slepton LOSP.  On the other hand if the LOSP decay arises from $L^\dagger L X'$ or $LEH X'$ then we expect decays $\tilde{l}^\pm_i \rightarrow l^\pm_j \tilde{x}'$ that reflect the flavor structure of the corresponding coupling matrix.  Thus the flavor structure will be a strong indication of whether the Higgs/Bino Portals are operative or not, since these theories are flavor blind.

\subsection{Signals from Re-Annihilations and Asymmetry}\label{adm}

The predicted relation for the FI mechanism in Table \ref{tab:introtab} corresponds to FI arising solely from LOSP decays.  In general, there will be FI contributions which arise from heavier superpartners, and in the cases of the Higgs and Bino Portals, the DM abundance from FI can be computed exactly, assuming the mass spectrum has been measured.  From this, the LOSP lifetime necessary to account for the measured DM abundance can be determined and checked against collider measurements.

More generally, this predicted relationship among measured quantities for FI, as well as those in the case of FO\&D shown in Table \ref{tab:introtab} will be verified if either of these are indeed the dominant mechanisms for DM production in our universe.  On the other hand, if these predicted relations are not satisfied, this would be an indication that a different mechanism of DM production is at work.  Depending on the values of $\tau$ and $\langle \sigma v\rangle$ measured, Figure \ref{fig:introplot} implies that the origin of DM may be FO\&D$_{\rm r}$, FI$_{\rm r}$, FO\&D$_{\rm a}$, or FI$_{\rm a}$.   Unfortunately, as shown in \cite{cosmopaper}, the re-annihilated variants, FO\&D$_{\rm r}$ and FI$_{\rm r}$ are not fully reconstructible because in these theories the final abundance of DM depends on $\langle \sigma v \rangle '$, the annihilation cross-section of $X'$, which is not accessible at the LHC.  Likewise, the asymmetric variants, FO\&D$_{\rm a}$ and FI$_{\rm a}$ both depend on a CP phase which may be very difficult to measure at the LHC.

Nonetheless, while these theories are usually not fully reconstructible, it is the case that their associated portal operators may still be probed at the LHC.  For example, just like FO\&D, FO\&D$_{\rm r}$ and FO\&D$_{\rm a}$ can only function if visible sector FO  yields an overabundance of $X$, and so the allowed LOSP candidates for these cosmologies are the bino-like neutralino and the right-handed slepton.  FO\&D$_{\rm r}$ and FO\&D$_{\rm a}$ work for a broad range of portal interactions, and the signatures are precisely the same as those for FO\&D shown in Tables \ref{tab:FOD1} and \ref{tab:FOD2} and in the first two rows of Tables \ref{tab:REsig} and \ref{tab:ROsig}.  

The range of LOSP lifetimes for FO\&D$_{\rm r}$ is large, and similar to that for FO\&D, as shown in Figure 11 of \cite{cosmopaper}.   On the other hand, FO\&D$_{\rm a}$ requires a long lifetime, reducing the number of LOSP decays that can be measured for the case of the neutral bino.  Furthermore,  as $\epsilon$ is reduced so FO\&D$_{\rm a}$ requires a larger over-abundance from LOSP FO, which will remove the charged slepton LOSP candidate.  FO\&D$_{\rm a}$ also requires the interference of two LOSP decay modes with differing hidden sector charges.  If the amplitudes for these two modes are very different the asymmetry is suppressed, hence, with a high luminosity both modes may be seen.  They could have the same or differing visible sector signatures, but they must have different hidden sector particles; for example $\tilde{x}'$ in one and $\tilde{y}'$, with larger mass, in the other.

What about FI? FI$_{\rm r}$ and FI$_{\rm a}$ are similar to FI in the sense that the LOSP can be any MSSM superpartner except for the bino.  Depending on the dominant operator, the LOSP decay signatures are as in Tables \ref{tab:REsig} and \ref{tab:ROsig}.   As shown in Figure 11 of \cite{cosmopaper}, a wide range of lifetimes are possible for both re-annihilated and asymmetric cases.  While the DM abundance from FI$_{\rm r}$ can never be reconstructed, since it depends on the annihilation cross section in the hidden sector, there may be special cases of FI$_{\rm a}$ that can be reconstructed.   As shown in  \cite{HallJMRWest}, FI$_{\rm a}$ results from decays of a non-LOSP and the leading contribution typically arises from the interference between a purely visible sector decay and a decay mode involving the hidden sector.  If one is lucky, the decay mode to the hidden sector may be dominated by a single interaction, and its coefficient learned from the LOSP lifetime.
If $B-L$ is to be created via FO\&D$_{\rm a}$, or via FI$_{\rm a}$ \cite{HallJMRWest}, the visible part of the portal operator must be R-parity odd so that the signals are those of Table \ref{tab:ROsig}.

The signals that arise in supersymmetric theories where a pre-existing asymmetry, in either baryons or DM, is shared by equilibration  via $R$-parity odd operators \cite{Chang:2009sv} have the same topologies as those shown in Table \ref{tab:ROsig}.  However equilibration of an asymmetry can be distinguished from the FO\&D$_{\rm a}$ and FI$_{\rm a}$ generation mechanisms as the LOSP requires a shorter lifetime that violates Eq. \ref{eq:tau}.






\section*{Acknowledgments}
L.H. thanks Karsten Jedamzik, John March-Russell and Stephen West for useful discussions.
This work was supported in part by the Director, Office of Science, 
Office of High Energy and Nuclear Physics, of the US Department of 
Energy under Contract DE-AC02-05CH11231and by the National Science 
Foundation on grant PHY-0457315.

\appendix

\section{Higgs Portal} \label{higgsportal-model}

In this appendix we present a more detailed discussion of the ``Higgs Portal'' introduced in Section \ref{sec:opanalysis}.  In particular, we will be concerned with UV motivations for the presence of the associated operator,
\begin{equation}\label{hp-1}
\lambda\,\int\,d^2\theta\,H_uH_dX' 
\end{equation}
where $\lambda$ is a dimensionless coupling.  Certainly, if we assign $X'$ to have even R-parity and R-charge $2-R_1$, then this is the leading supersymmetric operator connecting the visible and hidden sectors.  A priori, from an effective field theory point of view one would expect $\lambda \lesssim 1$.  However, as shown in \cite{cosmopaper}, the requirement that the FI mechanism give rise to the correct relic abundance fixes a lifetime for $X$ corresponding to $\lambda$ of order $10^{-12} - 10^{-11}$ for weak scale $X$ and $X'$ masses.  Thus, to generate a value for $\lambda$ which is sufficiently suppressed, we require additional model-building.

For example, this can be accomplished by separating the visible and hidden sectors along an extra dimension.  In this case $\lambda$ may be vanishing or exponentially small in the microscopic theory, but nonetheless generated in the IR after integrating out heavy ``connector'' fields bridging the visible and hidden sectors.  This can occur singlet kinetic mixing scenarios in which $X'$ kinetically mixes with a visible sector singlet, as in \cite{Cheung:2010jx}.  Alternatively, $\lambda$ may be small if $X'$ or the MSSM Higgs fields are composite operators, in which case $H_u H_d X'$ will arise from some primordial higher dimension operator suppressed by a high scale $M_*$.  In this case $\lambda$ will be proportional the compositeness scale over $M_*$, which may be quite small. 

However, it is also possible to naturally generate a suppressed $\lambda$ with supersymmetry breaking operators, assuming a different R-charge assignment of $X'$. This has the advantage of relying only on symmetry arguments. To this effect, we assign $X'$ to be R-parity even with  R-charge  equal to $-R_1$. We assume the existence of a spurion $\Phi =\theta^2 m_{3/2}$ which is the order parameter for supersymmetry breaking.  Since $\int d^4\theta H_u H_d X'$ vanishes by holomorphy, then the leading portal interactions are given by
\begin{eqnarray}\label{Lag}
{\cal L} &=& \frac{\kappa}{M_*}\int d^4\theta\,H_u\,H_d\,X'\,\Phi^\dagger + \frac{\tilde\kappa}{M_*}\int d^4\theta\,H_u\,H_d\,X'\,\Phi^\dagger \Phi\nonumber\\
 &= &\lambda\,\int d^2\theta\, H_u\,H_d\,X' + \tilde{\lambda}\,m_{3/2}\,h_u\,h_d\,x'
\end{eqnarray}
where $\kappa$ and $\tilde\kappa$ are coefficients of ${\cal O}(1)$, and $\lambda,\,\tilde{\lambda} = {\cal O}(1)\,\frac{m_{3/2}}{M_*}$. Thus, if $\lambda$ is effectively a supersymmetry breaking operator, then it is naturally suppressed for $M_* \gg m_{3/2}$, which is quite small.   Note that the Giudice-Masiero mechanism for $\mu$ and $B\mu$ can still operate if there is an additional spurion for supersymmetry breaking with the appropriate R-charge and R-symmetry.

From Eq.~(\ref{Lag}), the leading decays of all superpartners occur through the operator $H_u\,H_d\,X'$ with coefficient $\lambda$. Apart from visible sector couplings, masses and mixing angles (which we assume can be measured), all such processes only depend on the coefficient $\lambda$ and the mass of the LSP $m'$, as they only involve the operator in Eq.~(\ref{hp-1}). Thus, measuring ($\lambda,m'$) from the decays of the LOSP allows us to infer the decay rates for the heavier superpartners and reconstruct the entire FI contribution to the LSP abundance in the early Universe. Recall that as explained in Section \ref{sec:FI}, all 
processes $(\tilde{y},\tilde{x}\rightarrow \tilde{y}' + \textrm{SM})$ are assumed to be absent. This can easily occur if the R-parity and R-charge of $Y'$ forbid such couplings or if such processes are kinematically forbidden.   

Note that Eq.~(\ref{hp-1}) leads to a mass-mixing between the higgsino and $\tilde{x}'$ when one of the higgs fields gets a vev. This mass mixing allows the neutralino to mix with $\tilde{x}'$ via its higgsino component and decay to hidden sector particles.

\section{Bino Portal}
\label{app:bino}

In this appendix we present a brief exposition on the ``Bino Portal'' introduced in Section \ref{sec:opanalysis}.  The operator $B^\alpha$ automatically has an R-charge of one and is odd under R-parity. As a consequence, the hidden sector 
field that can couple to $B^\alpha$ must be a vector field with R-charge one and must be odd under 
R-parity. The leading gauge invariant operators are
\be 
\mathcal{L} = \lambda\, \int d^2 \theta \, B^\alpha X^\prime_\alpha  + \tilde\lambda\, \int d^2 \theta \, B^\alpha X^\prime_\alpha  \Phi
\ee
where $\lambda$ and $\tilde \lambda$ are dimensionless coupling constants and again $\Phi = \theta^2 m_{3/2}$ is the supersymmetry breaking spurion. The first term is supersymmetric and was studied in the gauge kinetic mixing scenario \cite{Holdom:1985ag}, which occurs when the hidden sector contains a $U(1)^\prime$ 
symmetry, with gauge field $X^\prime_\alpha$.  The second term is intrinsically supersymmetry breaking and will induce a mass mixing between the gauginos.
 Both terms will induce couplings between the hidden and visible sector states.  Also, note that the supersymmetric kinetic mixing operator, $\lambda B^{\alpha} X^\prime_\alpha$, 
is dimension four, so it may be generated from UV dynamics but is non-decoupling in the 
infrared. Also, since this Lagrangian does not violate any symmetries of the Standard Model, 
it is not too constrained by existing experiment.

The above portal operators contain a kinetic and mass mixing among gauginos which are gauge invariant, relevant operators given by
\be \label{eq:kmix}
\mathcal{L} 
\supset  i \lambda\,\tilde{b}\bar{\sigma}^{\mu} \partial_{\mu}\tilde{x}^\prime + \tilde \lambda \, m_{3/2}  \, \tilde b \tilde x' .
\ee 
In \cite{cosmopaper} it was shown that the requirement for the FI mechanism to yield 
the correct dark matter relic abundance fixes a LOSP lifetime which corresponds to $\lambda$ and $\tilde \lambda$ of order $10^{-12} - 10^{-11}$ for weak scale 
masses. The kinetic mixing term is generated from the 
underlying theory, and so to generate a sufficiently small $\lambda$ and $\tilde \lambda$ requires some constraints on the 
underlying theory.

One possibility is if the $U(1)^\prime$ is embedded in a GUT, then the there will be connector 
particles with couplings $g$ and  $g^\prime$ to $B^\alpha$ and 
$X^\prime_\alpha$ respectively. So, $B^\alpha$ and $X^\prime_\alpha$ will be connected by 
a loop of these heavy particles. The kinetic mixing term can be generated by integrating out 
the connector particles, which results in an induced coefficient for the kinetic mixing term in Eq.~(\ref{eq:kmix}):
\be 
\lambda &=& \sum_{i} \frac{g_i g_i^\prime}{16 \pi^2} \text{Log}
\left(\frac{\Lambda}{m_i} \right)
\ee
where $\Lambda$ is the UV cutoff and $m_i$ is the mass of the heavy particles in the loop.
In this case $\lambda$ may be made small if $g^\prime_i \ll g_i$. A more attractive possibility is if 
the underlying theory contained some non-Abelian group that was broken down to $U(1)^\prime$. In 
this case the $U(1)'$ gauge field is not primordial, so a  kinetic mixing term can only be generated with the help of a gauge symmetry breaking field $\Sigma$ which acquires
a VEV:
\be
 \int d^2 \theta B^\alpha X^\prime_\alpha \frac{\langle \Sigma \rangle}{M_{\star}} 
&\Rightarrow& \lambda = \frac{ \langle \Sigma \rangle }{M_{\star}} 
\ee 
giving rise to a naturally suppressed 
$\lambda$ if $\langle \Sigma \rangle \sim M_{\rm EW} \ll M_{\star}$.  This  mechanism will simultaneously suppress $\tilde \lambda$.

The gaugino kinetic mixing can 
be removed from the Lagrangian by a field shift of the gaugino component 
of $B_{\alpha}$ superfield:
$\tilde{B}_{\alpha} \rightarrow \tilde{B}_{\alpha} + \lambda\,\tilde{X}_{\alpha}^\prime$. However, this 
field redefinition induces hidden-visible sector couplings in the gauge-chiral 
interaction lagrangian:
\be 
\mathcal{L}_{\text{int}} &=& \int d^4 \theta 
\left(\Phi^\dagger_i e^{2 g_i B_\alpha} \Phi_i + \Phi^{\dagger \prime}_j 
e^{2 g^\prime_j X^{\prime}_\alpha} \Phi^\prime_j \right) \nonumber\\
&\rightarrow& \left(\Phi^\dagger_i e^{2 g_i (B_\alpha+\lambda\,\tilde{X}_{\alpha}^\prime)} \Phi_i + \Phi^{\dagger \prime}_j e^{2 g^\prime_j X^{\prime}_\alpha} \Phi^\prime_j \right) \nonumber\\
&\supset& g_i \phi^{\star}_i \psi_i \tilde{b} + \text{h.c}+g_i \lambda\,\phi^{\star}_i \psi_i \tilde{x}^\prime + \ldots 
\ee
where $\Phi_i$ and $\Phi^\prime_j$ collectively denote charged chiral superfields in the visible and hidden sectors, respectively.  Also $\phi_i$ and $\psi_i$ denote the scalar and fermionic components of $\Phi_i$, and so on. Thus, upon shifting 
$\tilde{b}$, a coupling $\lambda \tilde{x}^\prime \tilde{J}$ is generated where $\tilde{J}$ is the hypercharge 
supercurrent of the visible sector: $\tilde{J} = \sum_{\text{i=visible}} g_i \phi_i \psi_i$. 

In addition, since the bino is massive, the shift in $\tilde{b}$ generates a mass mixing 
between the bino and the hidden sector fermion $\tilde{x}^\prime$ : 
$m_{\tilde{b}} \tilde{b} \tilde{b} \rightarrow 
m_{\tilde{b}} \tilde{b}\tilde{b} + \lambda 
m_{\tilde{b}}\left( \tilde{b} \tilde{x}^{\prime} +\text{h.c} \right)$.  Moreover, the supersymmetry breaking operator, $\tilde\lambda m_{3/2} \tilde b \tilde x'$, contributes a direct mass mixing term.  After removing the gaugino kinetic mixing term and diagonalizing the gaugino mass matrix, the $\tilde b$ and $\tilde x'$ become maximally mixed since 
$m_{\tilde{b}} \sim m_{\tilde{x}^\prime} \sim M_{\rm EW}$.  Consequently, this gives rise to a coupling $\lambda \tilde{b} \tilde{J}^\prime$, with
$\tilde{J}^\prime = \sum_{\text{i=hidden}} g^\prime_i \phi^\prime_i \psi^\prime_i$. 
In summary the hidden-visible sector couplings induced by kinetic mixing are given schematically by
\[\mathcal{L}_{\text{int}} \approx 
\lambda \left(\tilde{x}^\prime \tilde{J} + \tilde{b} \tilde{J}^\prime \right) \]
where there is in actuality a relative $\mathcal{O}(1)$ mixing angle between these two terms fixed by the precise (weak scale) masses of the visible and hidden sector gauginos.  Consequently, given comparable visible and hidden sector gaugino masses, which is the case in our scenario of gravity mediated supersymmetry breaking, the total effect of the gauge kinetic mixing is to weakly couple the visible sector gaugino to the hidden sector supercurrent and vice versa.


\begin{thebibliography}{99}

\bibitem{Kolb:1988aj}
  E.~W.~Kolb and M.~S.~Turner,
  ``The Early Universe.'' (1988).

\bibitem{Baltz:2006fm}
  E.~A.~Baltz, M.~Battaglia, M.~E.~Peskin and T.~Wizansky,
  Phys.\ Rev.\  D {\bf 74}, 103521 (2006)
  [arXiv:hep-ph/0602187].


\bibitem{Cheung:2010gj}
  C.~Cheung, G.~Elor, L.~J.~Hall, P.~Kumar,
  JHEP {\bf 1103}, 042 (2011).
  [arXiv:1010.0022 [hep-ph]].



\bibitem{Hall:2009bx}
  L.~J.~Hall, K.~Jedamzik, J.~March-Russell and S.~M.~West,
  JHEP {\bf 1003}, 080 (2010)
  [arXiv:0911.1120 [hep-ph]].

\bibitem{Chang:2009sv}
  S.~Chang and M.~A.~Luty,
  arXiv:0906.5013 [hep-ph].


\bibitem{ADG}
N. Arkani-Hamed, A. Delgado and G.F. Giudice, arXiv:hep-ph/0601041.
\\
J.L. Feng, K.T. Matchev and F. Wilczek, hep-ph/0004043.

\bibitem{King:2006tf}
  S.~F.~King and J.~P.~Roberts,
  JHEP {\bf 0609}, 036 (2006)
  [arXiv:hep-ph/0603095].

\bibitem{Feng:2003uy}
  J.~L.~Feng, A.~Rajaraman and F.~Takayama,
  Phys.\ Rev.\  D {\bf 68}, 063504 (2003)
  [arXiv:hep-ph/0306024].

\bibitem{Feng:2004zu}
  J.~L.~Feng, S.~f.~Su and F.~Takayama,
  Phys.\ Rev.\  D {\bf 70}, 063514 (2004)
  [arXiv:hep-ph/0404198].


\bibitem{Covi:1999ty}
  L.~Covi, J.~E.~Kim and L.~Roszkowski,
  Phys.\ Rev.\ Lett.\  {\bf 82}, 4180 (1999)
  [arXiv:hep-ph/9905212].
\\
  L.~Covi, H.~B.~Kim, J.~E.~Kim and L.~Roszkowski,
  JHEP {\bf 0105}, 033 (2001)
  [arXiv:hep-ph/0101009].
\\
  T.~Asaka and T.~Yanagida,
  Phys.\ Lett.\  B {\bf 494}, 297 (2000)
  [arXiv:hep-ph/0006211].

\bibitem{Cheung:2010mc}
  C.~Cheung, Y.~Nomura and J.~Thaler,
  JHEP {\bf 1003}, 073 (2010)
  [arXiv:1002.1967 [hep-ph]].

\bibitem{Cheung:2010qf}
  C.~Cheung, J.~Mardon, Y.~Nomura and J.~Thaler,
  JHEP {\bf 1007}, 035 (2010)
  [arXiv:1004.4637 [hep-ph]].

\bibitem{ArkaniHamed:2006mb}
  N.~Arkani-Hamed, A.~Delgado and G.~F.~Giudice,
  Nucl.\ Phys.\  B {\bf 741}, 108 (2006)
  [arXiv:hep-ph/0601041].

\bibitem{Holdom:1985ag}
B.~Holdom,
Phys.\ Lett.\  B {\bf 166}, 196 (1986).

  K.~R.~Dienes, C.~F.~Kolda and J.~March-Russell,
  Nucl.\ Phys.\  B {\bf 492}, 104 (1997)
  [arXiv:hep-ph/9610479].

\bibitem{Glashow:1985ud}
S.~L.~Glashow,
Phys.\ Lett.\  B {\bf 167}, 35 (1986);
E.~D.~Carlson and S.~L.~Glashow,
Phys.\ Lett.\  B {\bf 193}, 168 (1987);
B.~Holdom,
Phys.\ Lett.\  B {\bf 259}, 329 (1991);
R.~Foot and X.-G.~He,
Phys.\ Lett.\  B {\bf 267}, 509 (1991);
K.~R.~Dienes, C.~Kolda and J.~March-Russell,
Nucl.\ Phys.\  B {\bf 492}, 104 (1997)
[arXiv:hep-ph/9610479].

\bibitem{Pospelov:2007mp}
M.~Pospelov, A.~Ritz and M.~Voloshin,
Phys.\ Lett.\  B {\bf 662}, 53 (2008)
[arXiv:0711.4866 [hep-ph]];
N.~Arkani-Hamed, D.~P.~Finkbeiner, T.~R.~Slatyer and N.~Weiner,
Phys.\ Rev.\  D {\bf 79}, 015014 (2009)
[arXiv:0810.0713 [hep-ph]];
M.~Pospelov,
Phys.\ Rev.\  D {\bf 80}, 095002 (2009)
[arXiv:0811.1030 [hep-ph]];
E.~J.~Chun and J.-C.~Park,
JCAP {\bf 0902}, 026 (2009)
[arXiv:0812.0308 [hep-ph]];
Y.~Bai and Z.~Han,
Phys.\ Rev.\ Lett.\  {\bf 103}, 051801 (2009)
[arXiv:0902.0006 [hep-ph]];
A.~Katz and R.~Sundrum,
JHEP {\bf 0906}, 003 (2009)
[arXiv:0902.3271 [hep-ph]];
B.~Batell, M.~Pospelov and A.~Ritz,
Phys.\ Rev.\  D {\bf 79}, 115008 (2009)
[arXiv:0903.0363 [hep-ph]];
R.~Essig, P.~Schuster and N.~Toro,
Phys.\ Rev.\  D {\bf 80}, 015003 (2009)
[arXiv:0903.3941 [hep-ph]];
M.~Reece and L.-T.~Wang,
JHEP {\bf 0907}, 051 (2009)
[arXiv:0904.1743 [hep-ph]];
D.~E.~Morrissey, D.~Poland and K.~M.~Zurek,
JHEP {\bf 0907}, 050 (2009)
[arXiv:0904.2567 [hep-ph]];
J.~D.~Bjorken, R.~Essig, P.~Schuster and N.~Toro,
Phys.\ Rev.\  D {\bf 80}, 075018 (2009)
[arXiv:0906.0580 [hep-ph]];
B.~Batell, M.~Pospelov and A.~Ritz,
Phys.\ Rev.\  D {\bf 80}, 095024 (2009)
[arXiv:0906.5614 [hep-ph]].
D.~P.~Finkbeiner and N.~Weiner,
  Phys.\ Rev.\  D {\bf 76}, 083519 (2007)
  [arXiv:astro-ph/0702587].


\bibitem{ArkaniHamed:2008qp}
N.~Arkani-Hamed and N.~Weiner,
JHEP {\bf 0812}, 104 (2008)
[arXiv:0810.0714 [hep-ph]].

\bibitem{Baumgart:2009tn}
M.~Baumgart, C.~Cheung, J.~T.~Ruderman, L.-T.~Wang and I.~Yavin,
JHEP {\bf 0904}, 014 (2009)
[arXiv:0901.0283 [hep-ph]].

\bibitem{Cheung:2009qd}
C.~Cheung, J.~T.~Ruderman, L.-T.~Wang and I.~Yavin,
Phys.\ Rev.\  D {\bf 80}, 035008 (2009)
[arXiv:0902.3246 [hep-ph]];

\bibitem{Cheung:2009su}
  C.~Cheung, J.~T.~Ruderman, L.~T.~Wang and I.~Yavin,
  JHEP {\bf 1004}, 116 (2010)
  [arXiv:0909.0290 [hep-ph]].

\bibitem{Walker:2009ei}
  D.~G.~E.~Walker,
  arXiv:0907.3142 [hep-ph].



\bibitem{ArkaniHamed:2004yi}
  N.~Arkani-Hamed, S.~Dimopoulos, G.~F.~Giudice and A.~Romanino,
  Nucl.\ Phys.\  B {\bf 709}, 3 (2005)
  [arXiv:hep-ph/0409232].

\bibitem{ArkaniHamed:2004fb}
  N.~Arkani-Hamed and S.~Dimopoulos,
  JHEP {\bf 0506}, 073 (2005)
  [arXiv:hep-th/0405159].

\bibitem{Meade:2010ji}
  P.~Meade, M.~Reece and D.~Shih,
  arXiv:1006.4575 [hep-ph].

\bibitem{CMS-stopped}
The CMS Collaboration,
{\tt http://cms-physics.web.\\cern.ch/cms-physics/public/EXO-09-001-pas.pdf}

\bibitem{Asai:2009ka}
  S.~Asai, K.~Hamaguchi and S.~Shirai,
  Phys.\ Rev.\ Lett.\  {\bf 103}, 141803 (2009)
  [arXiv:0902.3754 [hep-ph]].

\bibitem{Arvanitaki:2005nq}
  A.~Arvanitaki, S.~Dimopoulos, A.~Pierce, S.~Rajendran and J.~G.~Wacker,
  Phys.\ Rev.\  D {\bf 76}, 055007 (2007)
  [arXiv:hep-ph/0506242].


\bibitem{Hamaguchi:2004df}
K.~Hamaguchi, Y.~Kuno, T.~Nakaya and M.~M.~Nojiri,
Phys.\ Rev.\  D {\bf 70}, 115007 (2004)
[arXiv:hep-ph/0409248];
J.~L.~Feng and B.~T.~Smith,
Phys.\ Rev.\  D {\bf 71}, 015004 (2005)
[Erratum-ibid.\  D {\bf 71}, 0109904 (2005)]
[arXiv:hep-ph/0409278];
K.~Hamaguchi, M.~M.~Nojiri and A.~de Roeck,
JHEP {\bf 03}, 046 (2007)
[arXiv:hep-ph/0612060].


\bibitem{Cheung:2010jx}
  C.~Cheung and Y.~Nomura,
  arXiv:1008.5153 [hep-ph].

\bibitem{HallJMRWest}
 L.~J.~Hall, J.~March-Russell and S.~M.~West,
  arXiv:1010.0245 [hep-ph].


\end{thebibliography}
\end{document}